\documentclass{aastex}          
\usepackage{spr-astr-addons} 
\usepackage{epsfig,graphicx}
\usepackage[T2A]{fontenc}
\usepackage[cp1251]{inputenc}
\usepackage{amsmath,amssymb}
\usepackage{multirow}
\usepackage{graphicx}
\usepackage{epsfig}
\usepackage{subfigure}
\usepackage[english]{babel}
\usepackage{url}
\RequirePackage{color}
\usepackage[colorlinks]{hyperref}
\usepackage{ifpdf}

\begin{document}

\title{New results for electromagnetic quasinormal and quasibound modes of Kerr black holes}

\shorttitle{EM quasinormal and quasibound modes of Kerr black holes}
\shortauthors{Staicova D., Fiziev P.}

\author{Denitsa Staicova\altaffilmark{1}} 
\altaffiltext{1}{Institute for Nuclear Research and Nuclear Energy,
Bulgarian Academy of Sciences, bul. ``Tsarigrasko shose'' 72, Sofia 1784, Bulgaria}
\email{dstaicova@inrne.bas.bg}
\and 
\author{Plamen Fiziev\altaffilmark{2,3}}
\altaffiltext{1}{Sofia University Foundation for Theoretical and Computational Physics and
Astrophysics, 5 James Bourchier Blvd., 1164 Sofia, Bulgaria}
\altaffiltext{3}{JINR, Dubna, 141980 Moscow Region, Russia}
\email{fiziev@phys.uni-sofia.bg}

\begin{abstract}
The perturbations of the Kerr metric and the miracle of their exact solutions play a critical role in the comparison of predictions of general relativity with
astrophysical observations of compact massive objects. The differential equations governing the late-time ring-down of the perturbations of the Kerr metric, the Teukolsky Angular Equation and the Teukolsky Radial Equation, can be solved analytically in terms of confluent Heun functions. In this article, we solve numerically the spectral system formed by those exact solutions and we obtain the
electromagnetic (EM) spectra of the Kerr black hole.

Because of the novel direct way of imposing the boundary conditions, one is able to discern three different types of spectra: the well-known quasinormal modes (QNM), the symmetric with respect to the real axis quasibound modes (QBM) and a spurious spectrum who is radially unstable. This approach allows clearer justification of the term ``spurious'' spectrum, which may be important considering the
recent interest in the spectra of the electromagnetic counterparts of events producing gravitational waves.
\end{abstract}

\maketitle

\section{Quasi-normal modes of black holes}
During the long history of the study of the quasinormal modes (QNMs) of a black hole (BH) (\cite{RWE,ZRE,vish},\\ \cite{Teukolsky0},\cite{Teukolsky, Teukolsky2,Teukolsky-Kerr}, \cite{Teukolsky21, teukolsky, chan0, chan1,chan2}, \cite{QNM0}, \cite{QNM, det1,QNM2},\cite{Leaver, Leaver0, Q_N_M}, \\ \cite{high, special3}, \cite{special1, Fiziev1, QNM1}, \cite{Fiziev2, Fiziev3}, \\ \cite{BHB1},\cite{GW7}, \cite{QNM21}), the case of electromagnetic (EM) perturbations has been often ignored in favor to the gravitational one. It is considered that the gravitational output should be significantly more luminous than the EM one (\cite{GW3}), which combined with the strong absorption of the EM spectrum by the interstellar medium, makes the detection very difficult at the predicted low frequencies for the electromagnetic QNMs. On the other hand, the gravitational waves (GW) interact very weakly with matter and thus they can travel big distances without getting absorbed or scattered, i.e. without obscuring the signature of the body that emitted them. It is, therefore, reasonable to expect that the GW should be much better suited for studying the central engines of astrophysical events, such as gamma-ray bursts (GRBs).

The lack of GW detection from both LIGO and VIRGO (\cite{LIGO7,LIGO2,LIGO3,LIGO1,LIGO8,LIGO6} \footnote{Particularly puzzling being the lack of GW detection from short GRBs (\cite{LIGO9}) whose progenitors are expected to emit GWs in the range of sensitivity of the detectors.}), however, forced the introduction of the so-called multimessenger approach. With it, one can use the EM counterpart of the GW emission to gain more complete picture on the event and to improve the localization of the source and the sensitivity of the analysis of the GW data \cite{GW1,GW3, LIGO10, GW8, GW4}, \\ \cite{GW5}. 

In the case of short GRBs, the negative results may be explained by a GRB-generating process which in good approximation preserves the spherical symmetry of the central engine and thus admits only significant dipole radiation (EM waves) and no quadrupole one (GW waves). It can also be due to the unknown {\em model-dependent} ratio between EM and GW emission in more traditional processes. In the absence of GW detection, studying the characteristic EM spectrum of the object is the best path to understand the GRB central engine.

The discrete spectrum of complex frequencies called QNMs describe only the linearized perturbations of the metric. For this reason, they cannot describe the dynamics of the process during the early, highly intensive stage of the event, when the linearized theory is not applicable. In EM observations, however, we ``see'' only the tails of the event, being far from them. Furthermore, it is known from full numerical simulations that the QNMs dominate the late-time evolution of the object's response to perturbations (\cite{high}, \cite{headon}). 

The QNMs correspond to particular boundary conditions characteristic of the object in question. In the case of BHs,  the no-hair theorem states that they should depend only on the parameters of the metric, which means that measuring those frequencies observationally can be used to test the nature of the object -- a black hole or other compact massive object like super-spinars (naked singularities), neutron stars, black hole mimickers  etc. (\cite{special31,NB2,NB3,NB1,NB4,spectra}). It also can constrain additionally the no-hair theorem, which has recently been called into question for the case of black holes formed as a result of the collapse of rotating neutron stars \cite{GW6}.

An interesting possibility is to find a way to use the damping times of the EM quasinormal modes for comparison with observations. While the frequencies can be distorted by interaction with the surrounding matter, their damping times should be related to the variability-scale of the event. Recent simulations of jet propagation imply that the short-scale variability of the light-curve is due to the central engine and not to the interaction of the jet with the surrounding medium (see \cite{time-scale}). Such studies could be particularly interesting in the case of the central engines of GRBs, where numerical simulations although capable of describing {\em some} of the features of the GRBs light curves (for a review on GRBs, see \cite{GRB}), still struggle to explain the extended duration and characteristic time-variability (for a novel model working in this direction see for example \cite{Rezzolla2014}). 

Existing models of GRB central engine include a compact massive object (black hole or a milli-second magnetar) and extreme magnetic field ($\sim10^{15}G$ ) to accelerate and collimate the matter. Such central engine can be studied approximately by the linearized perturbations of the Kerr black hole. The spectrum of such perturbation do not depend on the origin of the perturbation, but only on the parameters of the compact massive object (mass and spin). In the idealized EM case, the perturbation is described by free EM waves in vacuum. While the astrophysical black holes are thought to be not charged, they are immersed into EM waves with different energy and origin. The black hole response to such EM perturbations in linear approximation will be, then, the QNM spectrum defined by the appropriate boundary conditions \footnote{Other conditions more suitable for describing a primary jet were studied in \cite{spectra}.}.

The choice of the rotating Kerr black hole, instead of the simpler Schwarzschild case, is due to the fact that rotation is critical in known energy extraction processes (for example resonant amplification or the Blandford-Znajek process \cite{teukolsky,BHB1,GW7}). Available observationally measured rotations of astrophysical compact massive objects show that there are many cases of near extremal values (for example the rotation rates of two astrophysical black holes, Sw J1644+57 with $a=0.9M$ and Sw J2058+05 with $a=0.99M$ \cite{rot}). 

To obtain the QNM spectra, one needs to solve two second-order linear differential equations: the Teukolsky radial equation (TRE) and  the Teukolsky angular equation (TAE) and to impose on them specific boundary conditions (\cite{QNM,QNM0}). Until recently, solving those equations analytically was considered impossible in terms of known functions, so approximations with  more simple wave functions were used instead.  The resulting system of spectral equations -- a connected problem with two complex spectral parameters: the frequency $\omega$ and the separation constant $E$ -- has been solved using different methods (\cite{special1,QNM1,special3,QNM21}) with notably the most often used being the method of continued fractions adapted by Leaver from the problem of the hydrogen molecule ion in quantum mechanics \cite{Leaver,Leaver0}. This method, while being successful in obtaining the QNMs spectra, has some specific numerical problems, for example in calculation of purely imaginary modes (\cite{Leaver0}, p.8) 
and also it doesn't account for the branch cuts in the exact solutions.

The analytical solutions of the TRE and the TAE can be written in terms of the confluent Heun function (for $a\neq M$) as done for the first time in \cite{Fiziev1,Fiziev3, Fiziev4, Fiziev2}. Those functions are the unique local Frobenius solutions of a second-order linear ordinary differential equation of the Fucsian type \cite{heun,heun1_,heun2_, heun3_, heun4_} with 2 regular singularities ($z=0,1$) and one irregular ($z=\infty$) and they are denoted as:
$\text{HeunC}(\alpha,\beta,\gamma,\delta,\eta,z)$ (normalized to $\text{HeunC}(\alpha,\beta,\gamma,\delta,\eta,0)=1$). 
The confluent Heun function as implemented in the software package \textsc{maple} was used successfully in our previous works \cite{Fiziev1,spectra,arxiv3,arxiv1}. The advantage of using the analytical solutions is that one can impose the boundary conditions on them {\em directly} (see \cite{Fiziev1,spectra}) and thus to be able to control all the details of the physics of the problem. It is thought that the Teukolsky-Starobinsky identities stem from the fact that both the TRE and TAE can be solved in terms of the confluent Heun function \cite{Teukolsky-Kerr}.

In a series of articles, we developed a method for solving numerically two-dimensional systems featuring the Heun functions (the two-dimensional generalization of the M\"uller method described in \cite{arxiv}) and we used it successfully in the case of gravitational perturbation ($s=-2$) of the Schwarzschild metric \cite{arxiv1}. Besides repeating with high precision the already known results, we used the epsilon-method (see below) to study the branch cuts in the spectral problem and to show that the $9^{th}$ mode for $s=-2$ is not purely imaginary, and thus cannot be algebraically special. Analysis of the potentials of the Regge-Wheeler equation (RWE) and the Zerilli equation (ZRE) showed that its properties are due to the existence of a branch cut on the imaginary axis for this mode \cite{special2,AS}. This result is directly obtainable from the actual solutions of the RWE and ZRE in terms of the confluent Heun functions.  

In this article, we continue the exploration of the application of the confluent Heun functions by studying the QNMs of the Kerr BH. Previous results, obtained in \cite{arxiv-Kerr} showed that using the confluent Heun function one can obtain the QNMs to a very good precision for the well-known lower modes. In the current work we are able to obtain the QNMs for a wide range of modes and rotational parameters and we show that there is a very good agreement between our results and those obtained with other methods. Additionally, we study the numerical stability \footnote{Here by numerical stability of a mode we will understand that small deviations in the parameters of the radial variable do not change the mode up to certain significant digits.  } of the solutions with respect to the position of the radial variable in the complex plane. Such a study cannot be carried out with the continued fractions method, where the radial variable does not enter explicitly into the equations.

Taking advantage of direct way of imposing the boundary conditions on the system, one obtains not only the QNM spectrum but also the quasibound (quasiresonant) one (QBM). The QBMs form a discrete spectrum of frequencies with negative imaginary parts, obtained by imposing boundary conditions inverse to those of the QNMs. Those boundary conditions can correspond to time-inversion and are expected to be part of the solution due to its symmetries. They can also be considered as unstable bounded waves reflected by the horizon and infinity. An example of the study of the quasibound states in the case of massive vector propagating on the Schwarzschild space-time can be found in \cite{QBM}. Finally, we obtain an additional spectrum which is found to be spurious with respect to the so-defined spectral problem.

\section{The Teukolsky angular equations}
In Chandrasekhar's notation, the Teukolsky Master Equation (\cite{teukolsky}), for $|s|=1$, is separable under the substitution $\Psi\!=\!e^{i(\omega t\!+\!m\phi)}S(\theta)R(r)$, where $m=0, \pm 1, \pm 2$ for integer spins and $\omega$ is the complex frequency. Due to the choice of this form of $\Psi$, the sign of $\omega$ differs from the one Teukolsky used, and the stability condition, guaranteeing that the perturbations will damp with time reads $\Im(\omega)>0$.

The TAE for EM perturbations ($s=-1$) has 16 classes of exact solutions $S(\theta)$ in terms of the confluent Heun functions (for full details see \cite{Fiziev3}). To fix the spectrum approximately, one requires an additional regularity condition for the angular part of the perturbation, which means that if we choose one solution, $S_1(\theta)$, regular around the one pole of the sphere ($\theta=0$) and another, $S_2(\theta)$, which is regular around the other pole ($\theta=\pi$), then in order to ensure a simultaneous regularity, the Wronskian of the two solutions should become equal to zero, $W[S_1(\theta),S_2(\theta)]=0$. This gives us one of the equations for the two-dimensional spectral system defining QNM.

In \cite{Fiziev3}, there are four pairs of Wronskians, each pair being valid in a sector of the plane $\{s,m\}$. 
The Wronskians we used to obtain the spectra are:
\begin{align}
 W[S_1,S_2]=\frac{\text{HeunC}'(\alpha_1,\beta_1,\gamma_1,\delta_1,\eta_1,\left( \cos \left( \pi/6  \right)  \right) ^{2})}{\text{HeunC}(\alpha_1,\beta_1,\gamma_1,\delta_1,\eta_1,\left( \cos \left( \pi/6  \right)  \right) ^{2})}+\notag\\
\frac{\text{HeunC}'(\alpha_2,\beta_2,\gamma_2,\delta_2,\eta_2,\left( \sin \left( \pi/6  \right)  \right) ^{2})}{\text{HeunC}(\alpha_2,\beta_2,\gamma_2,\delta_2,\eta_2,\left( \sin \left( \pi/6  \right)  \right) ^{2})}+ p=0
\label{Wr1}
\end{align}
\noindent where the derivatives are with respect to $z$ and the values of the parameters for the two confluent Heun functions for each $m$ are as follows:

For the case $m=0$:
$\alpha_1= 4\,a\omega,
\beta_1= 1,
\gamma_1=- 1,
\delta_1=4\,a\omega,
\eta_1=1/2-E-2\,a\omega-{a}^{2}{\omega}^{2}$ and

$\alpha_2=-4\,a\omega,
\beta_2=1,
\gamma_2=1,
\delta_2=-4\,a\omega,
\eta_2=,1/2-E+2\,a\omega-{a}^{2}{\omega}^{2}$, $p=\frac{1}{\left( \sin \left( \pi/6  \right)  \right) ^{2}}$

For the case $m=1$:
$\alpha_1=-4\,a\omega,
\beta_1=2,	
\gamma_1=0,
\delta_1=4\,a\omega,
\eta_1=1-E-2\,a\omega-{a}^{2}{\omega}^{2}$
and

$\alpha_2= -4\,a\omega,
\beta_2=0,
\gamma_2=2,
\delta_2=-4\,a\omega,,
\eta_2=1-E+2\,a
\omega-{a}^{2}{\omega}^{2}$ and $p=-4\,a\omega$

For the case m=2:
$\alpha_1= -4\,a\omega,
\beta_1=3,
\gamma_1=-1,
\delta_1=4\,a\omega,
\eta_1=5/2-E-2\,a\omega-{a}^{2}{\omega}^{2}$
and

$\alpha_2= -4\,a\omega,
\beta_2=1,
\gamma_2=-3,
\delta_2=-4\,a\omega,
\eta_2=5/2-E+2\,a\omega-{a}^{2}{
\omega}^{2}$ and $p=8-4a\omega.$

\noindent where we use $\theta=\pi/3$ (the QNMs should be independent of the choice of $\theta$ in the spectral conditions).

This form of the Wronskians, different from the one in \cite{Fiziev3}, is chosen to improve the numerical convergence of the root-finding algorithm. 

\section{The Teukolsky radial equation}
The TRE differential equation is of the confluent Heun type, with $r=r_{\pm}$ regular singular points and $r=\infty$ -- irregular one. As it was noted in \cite{spectra}, the point $r=0, \theta=\pi/2$ is not a singularity for this equation. 
The solutions of the TRE for $r>r_{+}$, are :
\begin{align}
&R(r)\!=\!C_1R_1(r)+C_2R_2(r), \text{for} \label{R}\\
&R_1(r)\!=\!e^{\frac{\alpha\,z}{2}}(r\!-\!r_+)^{\frac{\beta\!+\!1}{2}}(r\!-\!r_-)^{\frac{\gamma\!+\!1}{2}}\text{HeunC}(\alpha,\beta,\gamma,\delta,\eta,z)\!\notag\\
&R_2(r)\!= R_1(r) (\beta\to-\beta) \notag
\end{align}
\noindent where $z=-\frac{r-r_+}{r_+-r_-}$ and the parameters are:

\begin{footnotesize}
\begin{align*}
\alpha &\!=\!-2\,i \left( {\it r_{_+}}\!-\!{\it r_{_-}} \right) \omega,  
\beta \!=\!-\!{\frac {2\,i(\omega\,({a}^{2}+{{\it r_{_+}}}^{2})\!+\!am)}{{\it r_{_+}}\!-\!{\it r_{_-}}}}\!-\!1,\\
 \gamma &\!=\!{\frac {2\,i(\omega\,({a}^{2}\!+\!{{\it r_{_-}}}^{2})\!+\!am)}{{\it r_{_+}}\!-\!{\it r_{_-}}}}\!-\!1, 
\delta \!=\!2i\!\left({\it r_{_-}}\!-\!{\it r_{_+}}\right)\!\left(\omega\!-\!i \left( {\it r_{_-}}\!+\!{\it r_{_+}} \right) \omega^2 \right)\!
\end{align*}
\end{footnotesize}

\begin{footnotesize}
\begin{align*}
\eta &=\!\frac{1}{2}\frac{1}{{ \left({\it r_{_+}}\!-\!{\it r_{_-}}\right) ^{2}}} \Big[ 4{\omega}^{2}{{\it r_{_+}}}^{4}\!+\! 4\left(i \omega
\!-\!2{\omega}^{2}{\it r_{_-}}\right) {{\it r_{_+}}}^{3}\! +\! (1\!-\!4a\omega\,m\!-\\
&\!2{\omega}^{2}{a}^{2}\!-\!2E ) \left( {{\it r_{_+}}}^{2}\!+\!{{\it r_{_-}}}^{2}\right) \! + 4\left(i\omega\,{\it r_{_-}} \!-\!2i\omega\,{\it r_{_+}}\!+\!E\!-\!{\omega}^
{2}{a}^{2}\!-\!\frac{1}{2} \right)\times \\
&{\it r_{_-}}\,{\it r_{_+}}\!-4{a}^{2} \left(m\!+\!\omega\,a \right) ^{2} \Big].
\end{align*}
\end{footnotesize}
Here the general solution is constructed following the theory of the Fucsian equations. Accounting for the symmetries of the confluent Heun function, the solutions \eqref{R} coincide with those in 	\cite{Fiziev3}(with $\omega\!\to \!-\!\omega$). \footnote{It is important to emphasize that the so obtained solutions cannot be used for extremal KBH ($a=M$) since in this case, the differential equation is of the biconfluent Heun type.}

The TRE has 3 singular points $r_-,r_+,\infty$ and in order to fix the spectrum, one needs to impose specific boundary conditions on two of those singularities (i.e. to solve the central two-point connection problem
\cite{heun3_}). Different boundary conditions on different pairs of singular points will mean different physics of the problem. The QNM spectrum is obtained trough the so called black hole boundary conditions (BHBC) -- waves going simultaneously into the event horizon ($r_+$) and into infinity -- following the same reasoning as in \cite{spectra} where additional details can be found. Then, the BHBC read:
\begin{enumerate}
 \item BHBC on the KBH event horizon $r_+$.

For $r\to r_+$, from $r(t) = r_+ +e^{\frac{-\Re(\omega)t+const}{\Im(n_{1, 2})}}\to r_{+}$, where $n_{1,2}$ are the powers of the factor $(r-r_+)^{n_{1,2}}$ in $R_{1,2}$. Thus for $m=0$, the only valid solution in the whole interval $\Re(\omega)\in (-\infty,\infty)$ is $R_2$, while for $m\neq0$, the solution $R_2$ is valid for $\Re(\omega) \not\in (-\frac{ma}{2Mr_+},0)$, in the rest of the interval one needs to use $R_1$ to describe ingoing waves. When this condition is not fulfilled, the spectrum corresponds to waves going out of the horizon -- a white hole case. 

\item BHBC at infinity.

At $r\to \infty$, the solution is a linear combination of an ingoing ($R_{\leftarrow}$)
and an outgoing ($R_{\rightarrow}$) wave: $R=C_{\leftarrow}\,R_{\leftarrow}+C_{\rightarrow}\,R_{\rightarrow},$
where $C_{\leftarrow}$, $C_{\rightarrow}$ are unknown constants and $R_{\leftarrow}, R_{\rightarrow}$ are found using the asymptotics of the confluent Heun function as in \cite{heun3_,Fiziev3}.

To ensure only outgoing waves at infinity, one needs to have $C_{\leftarrow}=0$.

To achieve this,  one first finds the direction of the steepest descent in the complex plane $\mathbb{C}_r$ for which  $\lim\limits_{r\to\infty}\frac{R_{\rightarrow}}{R_{\leftarrow}}=r^{-4i\,\omega\,M+2} e^{-2i\omega\,r}= 0$ tends to zero the most quickly: $\sin(\arg(\omega)\!+\!\arg(r))\!=\!-\!1$. This gives us the relation $r=\mid\! r\! \mid e^{3/2i\pi-i\,\arg(\omega)}$ (\cite{Fiziev1}), between $\omega$ and $r$, which is exact only if one uses the first term of the asymptotic series for the confluent Heun function (i.e. $\text{HeunC} \sim 1$). 

To completely specify the spectrum $\{\omega_{n,m}, E_{n,m}\}$, for $R=R_2$ ($R_1$ is analogous), it is enough to solve:
\begin{align}
C_{\leftarrow}\!=\!r^{2\!+\!i\,\omega\!+\!\frac{2i\,m\,a\!+\!i\,\omega}{r_+\!-\!r_-}}\text{HeunC}(\alpha,\!\!-\!\beta,\gamma,\delta,\eta,z)\!=\!0,\hspace{-6.5px}
\label{Rbc}
\end{align}
\end{enumerate}
The inverse boundary conditions, namely waves outgoing from the horizon and outgoing from infinity ($\lim\limits_{r\to\infty}\frac{R_{\leftarrow}}{R_{\rightarrow}}=0)$ give the QBM spectrum. Although the QBM frequencies are considered unphysical, because they lead to non-damping waves and thus to a black hole bomb, mathematically the differential system describes those states on equal footing as the QNMs. 

\section{The branch cuts of the radial solution}
Equation \eqref{Rbc} relies on the direction of the steepest descent defined by the phase condition $\sin(\arg(\omega)+\arg(r))  \lessgtr 0$ (upper sign for QNMs, lower -- for QBMs). This approximate direction was chosen ignoring the higher terms in the asymptotic expansion of the solution around the infinity point. Therefore, one can expect that the true path in the complex plane may not be a straight line but a curve. In principle, the spectrum should not depend on this curve as long as $r$ stays in the sector of the complex plane where $\lim\limits_{r\to\infty}\frac{R_{\rightarrow}}{R_{\leftarrow}}= 0$ for QNM (or the inverse for the QBM). 

Complications may arise due to the appearance of branch cuts (BC) in the confluent Heun function. In this numerical realization, as a branch cut is chosen the semi-infinite interval $z\in(1,\infty)$ on the real axis. 

Because of the phase condition, the branch cuts in the complex $r$-plane, appear also in the complex $\omega$-plane. In order to localize those branch cuts, we utilize the $\epsilon$-method  (\cite{arxiv3,arxiv1}) -- we add a small parameter $\epsilon<<1$ to the phase condition $\pi<\arg(r)\!+\!\arg(\omega)+\epsilon\!<2\pi$ (for QNMs), thus effectively moving $r=|r|e^{i\arg(r)}$ in the complex plane. 

Then the observed branch cuts are as follow:
\begin{enumerate}
 \item For $r$--real, one encounters a BC of the confluent Heun function. It is a line with equation: $\Im(\omega)/\Re(\omega)=\tan(3/2\pi+\epsilon\pi/2)=-\cot(\epsilon \pi/2)$, which rotates when $\epsilon$ changes.
 \item If $\Im(\omega)=0$, then one encounters the BC of the argument-function, lying on the negative real axis $\Re(\omega)\in(-\infty,0)$. This BC is important only when $a\to M$ when the frequencies are almost real.
 \item If $\Re(\omega)=0$ and $\Im(\omega)=2n$ , $n=1,2,3..$, then one can have $\Im(r)=0$ for certain values of $\epsilon$ and thus to reach the BC of the confluent Heun function on the real axis. This condition can affect almost imaginary modes such as the algebraically special modes.
\end{enumerate}

The knowledge of the branch cuts is critical for the successful numerical evaluation and analysis of the results. Because the phase condition introduces additional complexity in this respect, we prefer to vary $\arg(r)$ in $r=|r|e^{i\arg(r)}$ directly and afterwards to check whether the QNM or QBM phase condition is satisfied.

\section{Numerical algorithms}
The spectral equations we need to solve to find the spectrum $\omega_{n,m}(a)$ for $M=1/2$ are Eqs.\eqref{Wr1} and \eqref{Rbc}. This system represents a two-dimensional connected problem of two complex variables -- the frequency $\omega$ and the separation parameter $E$ -- and in both of its equations one encounters the confluent Heun function and in the case of the TAE -- their derivatives.

Because conventional methods like the Newton method and the Broyden method do not handle well such systems, our team developed a new method, the two-dimensional M\"uller algorithm, which proved to be  more suitable for the problem (for details see \cite{arxiv3,arxiv, arxiv1}). It relies on the M\"uller method, which is a quadratic generalization of the secant method having better convergence than the latter. The new algorithm does not need the evaluation of derivatives, thus saving time and avoiding some known difficulties in their evaluation outside the $|z|<1$ circle. It is important to note that {\bf both $\omega,E$ are found directly from the spectral system (Eqs. \eqref{Wr1} and \eqref{Rbc}) and with equal precision}. The results are obtained with \textsc{maple 17} code with software floating point number set to 64 and precision of the root-finding algorithm set to 25 digits. 

\section{Numerical results for electromagnetic QNMs}
To check the precision of the method we use as control frequencies $\{\omega_{n,m}^B,E_{n,m}^B\}$ the numbers published by Berti et al. (\cite{special31,special3}), which can be found on: \small \url{http://www.phy.olemiss.edu/~berti/qnms.html}. \normalsize \\ They were obtained using the continued fractions method, which is still considered as the most accurate method for calculation of the QNMs from the KBH. The available control frequencies are $n=0..6$ for $l=1$ and $n=0..3$ for $l=2$.

We use $|r|=110$ as the actual numerical infinity and $M=1/2$ as the mass of the object.

\subsection{Non-rotating BH}

It is well known that when there is no rotation ($a=0$), the electromagnetic QNMs come in pairs symmetrical to the imaginary axis $\omega_{n,m}^\pm=\pm|\Re(\omega_{n,m})|+i\Im(\omega_{n,m})$ ($n=0,1..$ numbering the mode). In this case, the system reduces to one equation -- the radial function \eqref{Rbc} (for $E=l(l+1),l=1,2..$), solved here using the one-dimensional M\"uller algorithm.

In our numerical studies, we vary $\arg(r)$ directly to find the zeros of the two equations $R_1(r)$ and $R_2(r)$ and then we classify the modes depending on the boundary conditions they satisfy -- QNM, QBM, others.  Thus, we have frequencies in the 4 quadrants of the complex plane, i.e. : $\omega_{n,m}=\pm|\Re(\omega_{n,m})|\pm i\Im(\omega_{n,m})$ ($n=0,1..$).

\begin{figure}[!h]
\vspace{-0cm}
\centering
\subfigure[\, m=0, l=1, $\omega$]{\includegraphics[width=117px,height=110px]{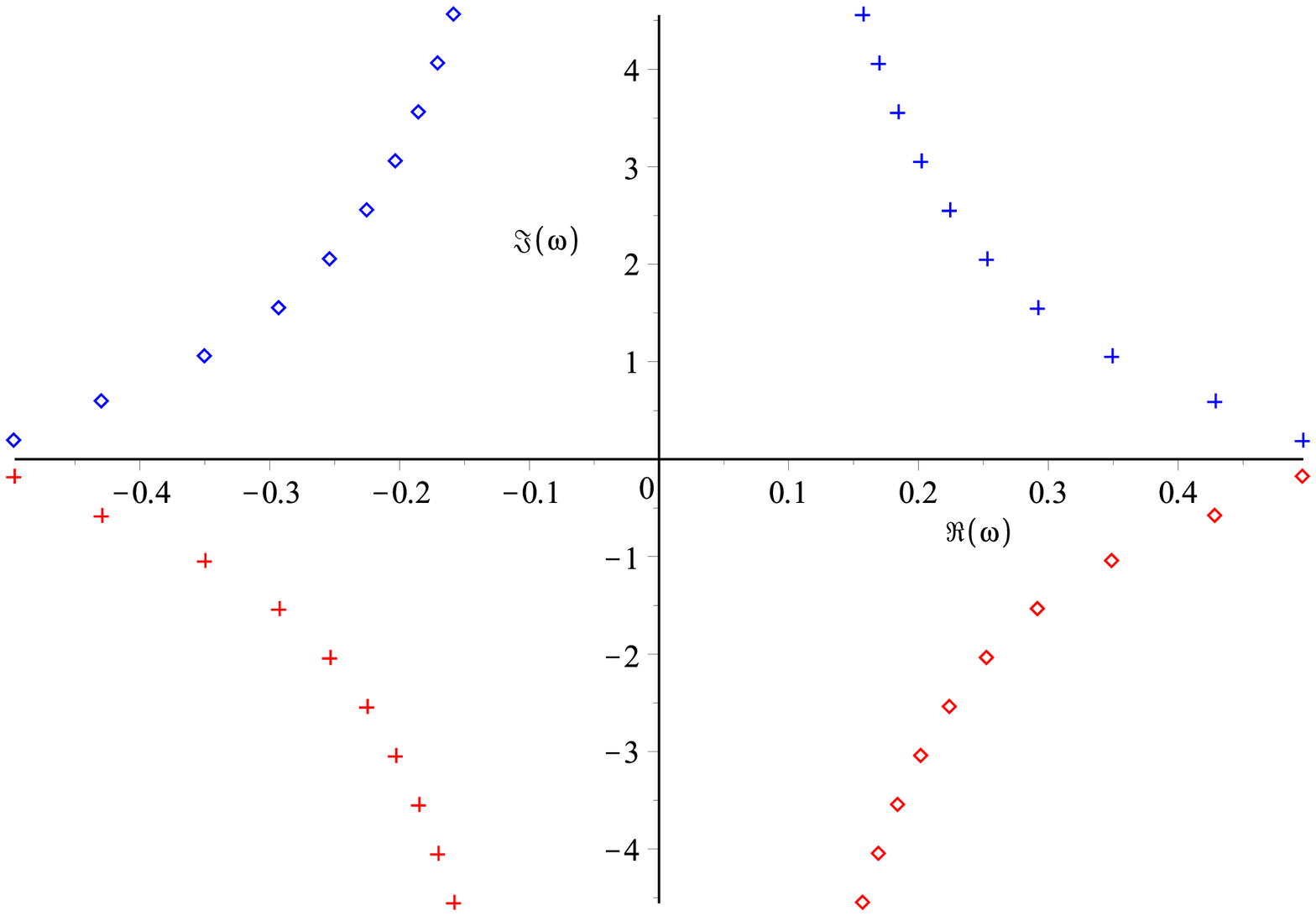}}
\subfigure[\,  m=0, l=1, $\sin(\arg(\omega)+\arg(r))$]{\includegraphics[width=117px,height=110px]{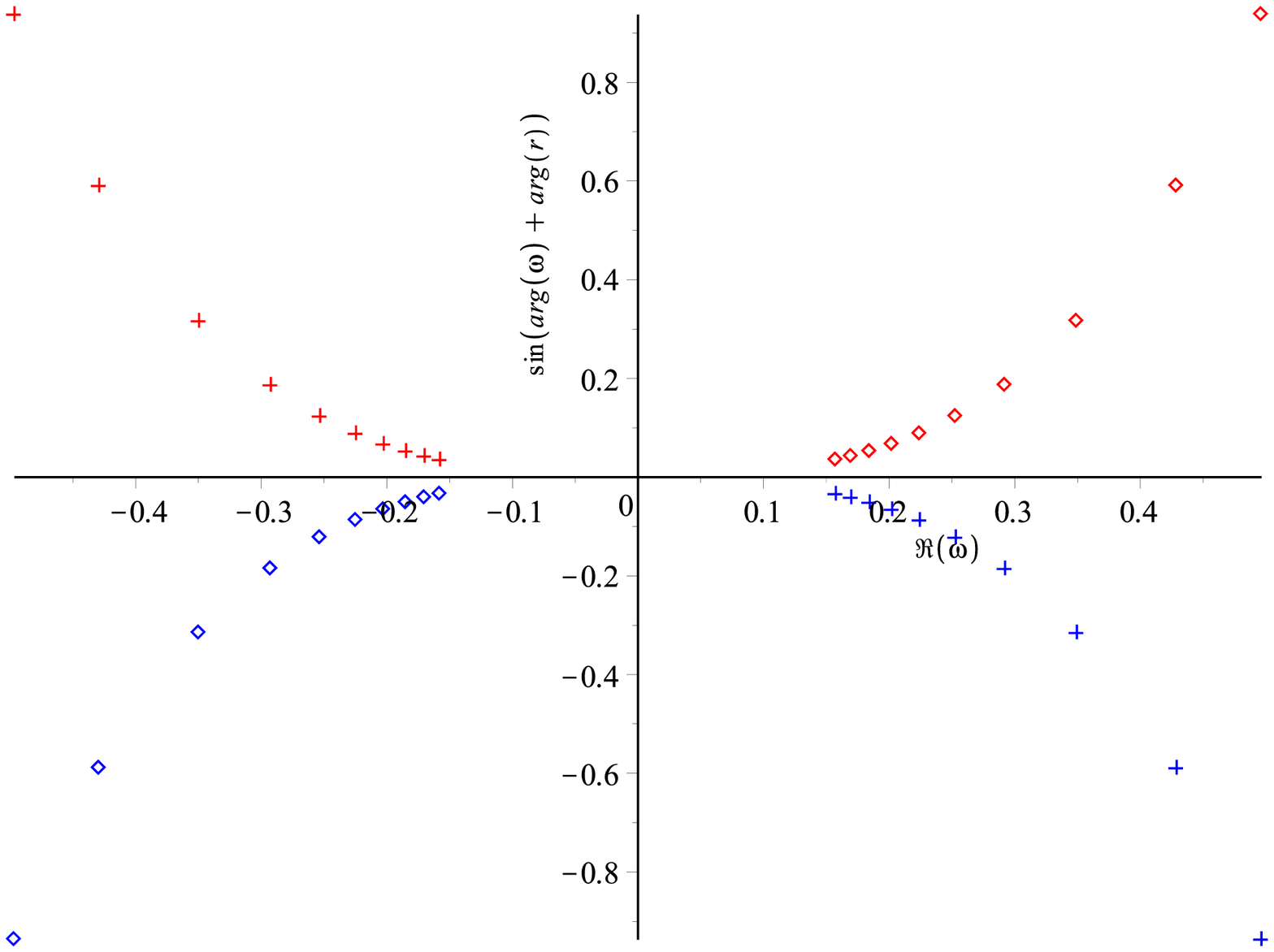}}
\caption{(a) The complex frequencies for $a=0$, for $m=0, l=1$. (b) the boundary condition $\sin(\arg(\omega)+\arg(r))$ for them. The red diamonds are obtained from $R_1(r)$ with $\arg(r)=1/2\pi$, the red crosses -- from $R_1(r)$ with $\arg(r)=3/2\pi$, the blue diamonds -- from $R_2(r)$ with $\arg(r)=1/2\pi$, the blue crosses -- from $R_2(r)$ with $\arg(r)=3/2\pi$. }
\label{m0_0}
\end{figure}

The results for $m=0, l=1$ can be seen on Fig. \ref{m0_0}: on a) the complex frequencies, on b) the corresponding boundary conditions at infinity ( $\sin(\arg(\omega)\!+\!\arg(r))\! \lessgtr\!0$). There is a clear symmetry with respect to both the real and the imaginary axes. Also, from the boundary conditions, one can easily differ between the two types of modes -- QNM and QBM. 

A numerical comparison of QNMs with the frequencies obtained by Berti et al. shows that the average deviation is $|\omega_{n,m}^B-\omega_{n,m}| \approx 10^{-10}$.

We checked the dependence of QNM/QBM modes on $|r|$ and, as expected, the modes are stable with respect to an increase in $|r|$, which means that $|r|=110$ is a valid actual infinity. It also confirms that in both cases, $\omega\nsim r$. This requirement comes from the ansatz in the separation of variables from which one obtains the TRE and the TAE. 

Another stability test is examining the dependence $\omega(\epsilon)$ (or $\arg(r)$). The so-found QNM and QBM are stable with 20 digits in certain intervals of $\epsilon$, whose width depends on $n$. 
Such stability intervals are related to the direction of steepest descent, as explained above. 

\subsection{Rotating KBH}

The results for $a=[0..M)$ can be seen on figures $2-7$.

For the QNM modes, when $a\neq0$, the symmetry with respect to the imaginary axis $\omega_{m,n}^{\pm} =\pm|\Re(\omega_{m,n})|+i\Im(\omega_{m,n})$ breaks down, but it is replaced by the symmetry:
 $$\{\Re(\omega_{m,n}^\pm),\Im(E_{m,n}^\pm),m\}\to \{\!-\!\Re(\omega_{m,n}^\pm),\!-\!\Im(E_{m,n}^\pm),\!-\!m\},$$
Thus to study the behavior of the modes for $a\in[0,M)$ for all $m$, it is enough to trace both $\omega_{m,n}^{\pm}(a)$ for only $m>0$ (the index $l$ here is omitted to simplify notations).

\begin{figure}[!htb]
\centering
\subfigure[$\omega_{0,3}(a)$]{\includegraphics[width=117px,height=110px]{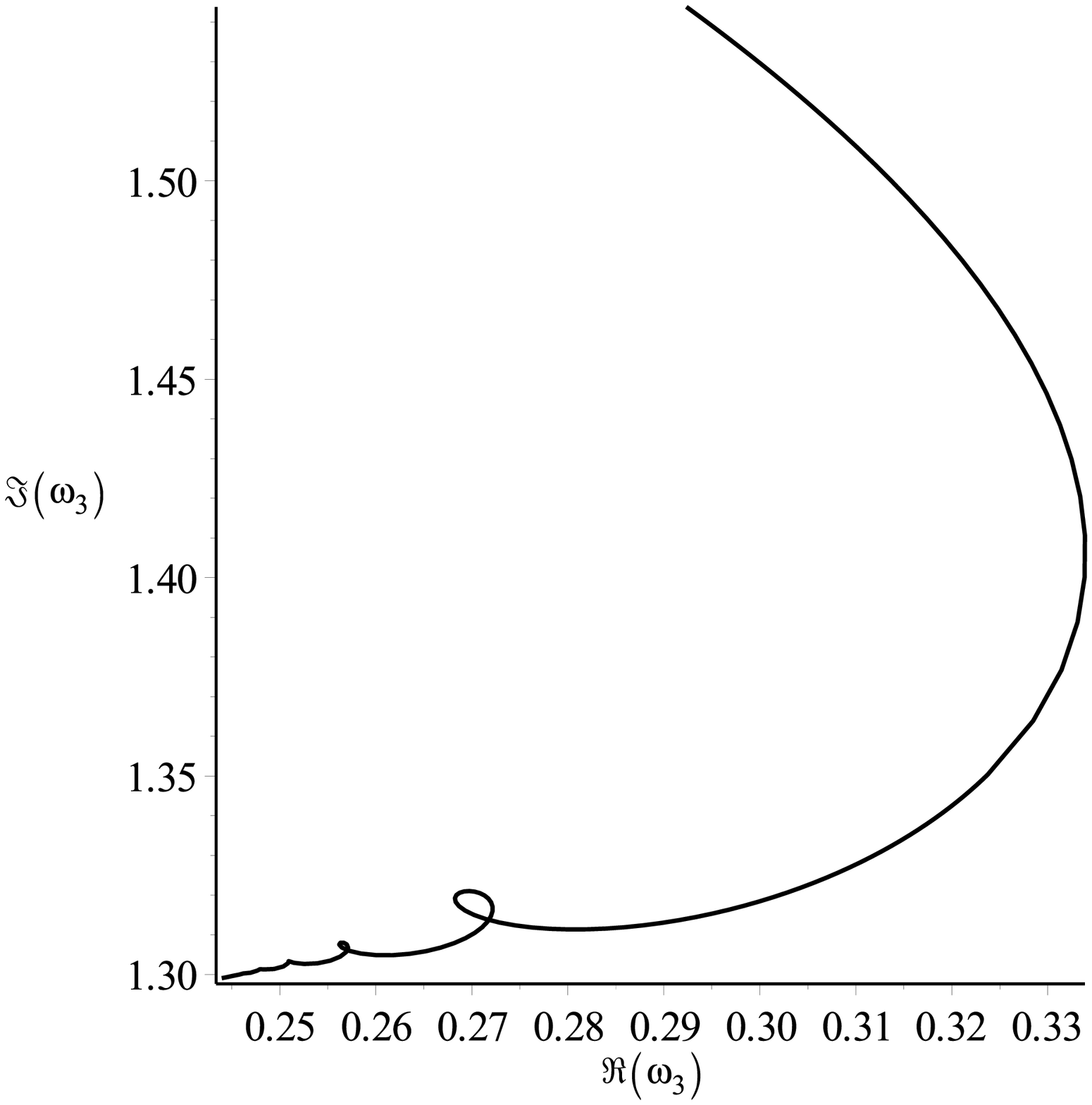}}
\subfigure[$E_{0,3}(a)$]{\includegraphics[width=117px,height=110px]{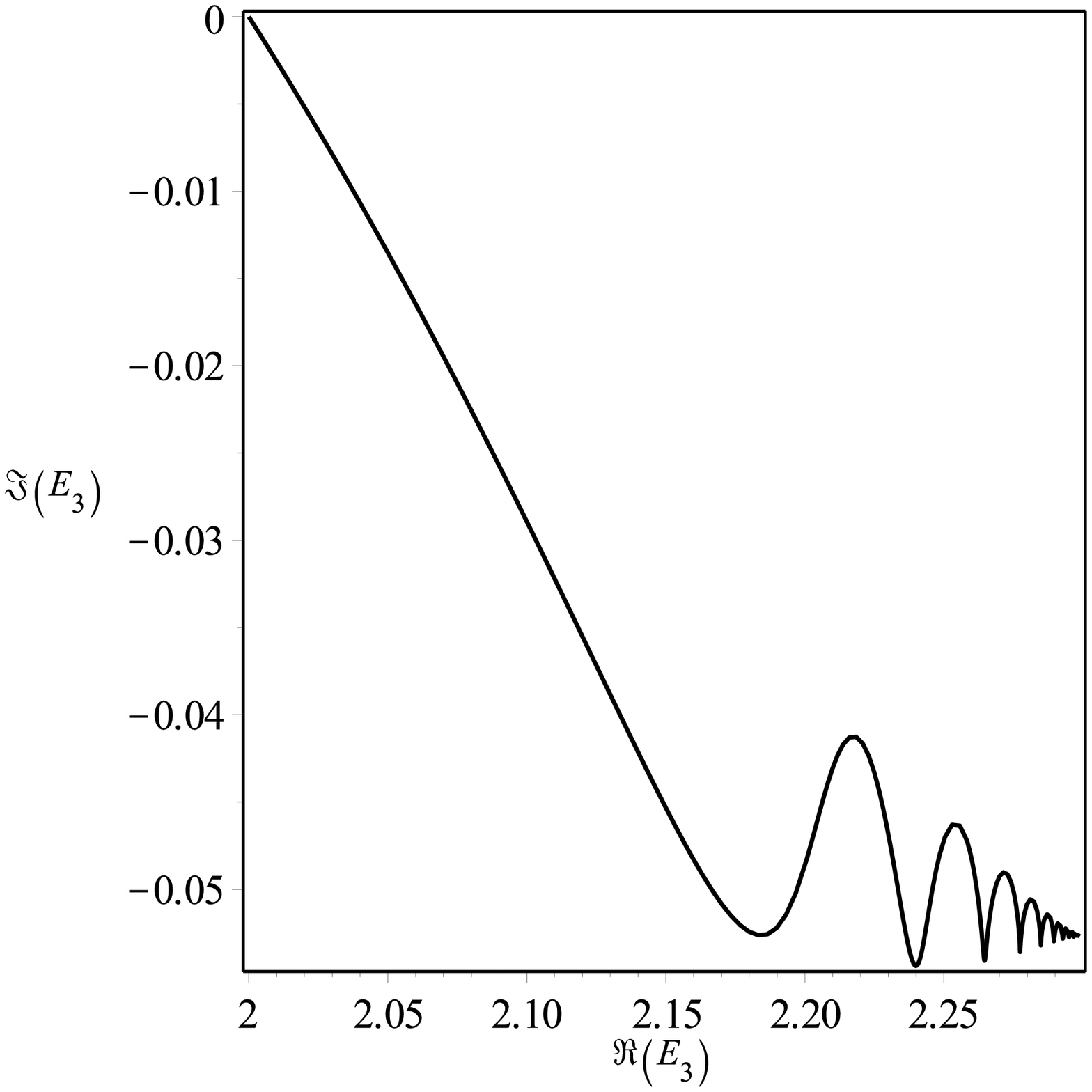}}
\caption{A complex plot of $\omega_{0,3}(a)$ and  $E_{0,3}(a)$. }
\label{n3}
\end{figure}

If one considers both the QNM and the QBM, i.e. the roots of the transcendental system in all the 4 quadrants of the complex plane  (${I},{II},{III},{IV})$ for $m=0, 1$, one observes the symmetry: \begin{align*}
&\Re(\omega_{I})=\Re(\omega_{IV}), \Re(\omega_{II})=\Re(\omega_{III}),\\
&\Im(\omega_{I})=-\Im(\omega_{II}) ,\Im(\omega_{III})=-\Im(\omega_{IV})
\end{align*}
(and analogously for $E$). This symmetry is preserved at least up to $n<4$ within the precision of the numerical method. 

\begin{figure}[!htb]
\vspace{-0cm}
\centering
\subfigure{\includegraphics[width=117px,height=110px]{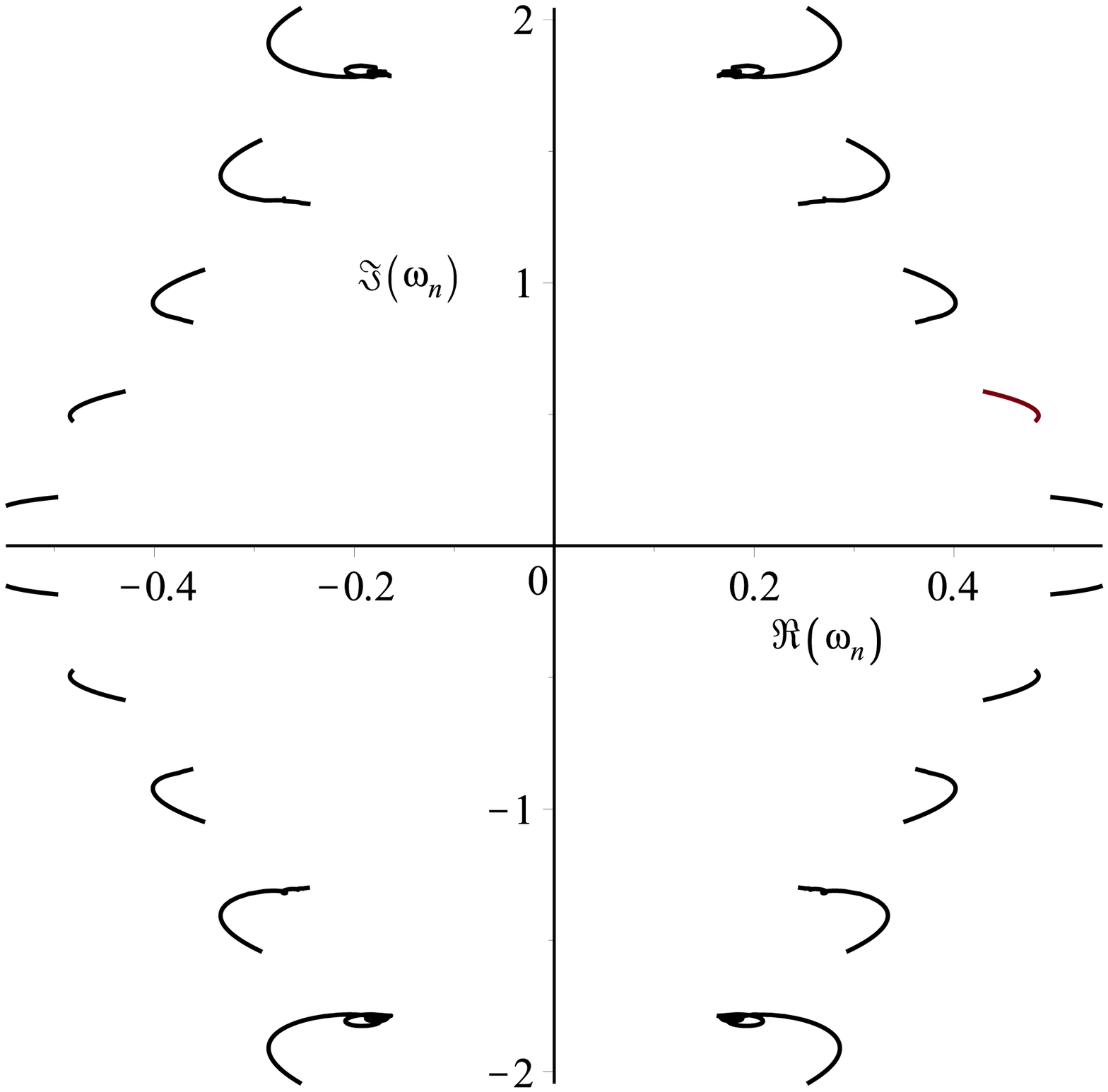}}
\subfigure{\includegraphics[width=117px,height=110px]{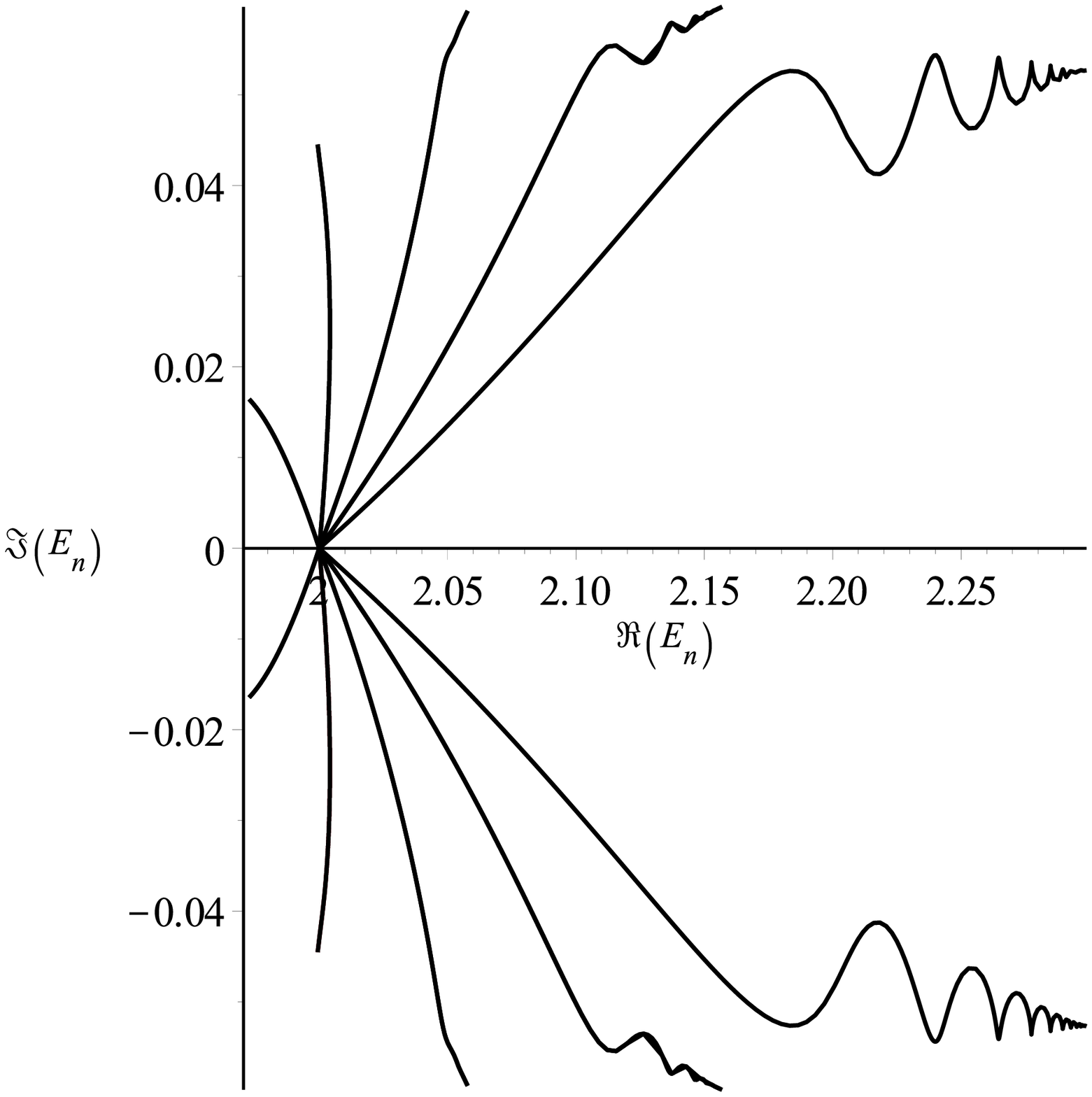}}
\caption{Complex plots of $\omega_{0,n}(a)$ and $E_{0,n}(a)$ for $a=[0,M)$ the first 5 modes with both positive and negative real parts}
\label{m0_all_o}
\end{figure}

\begin{figure}[!htbp]
\centering
\subfigure{\includegraphics[width=117px,height=110px]{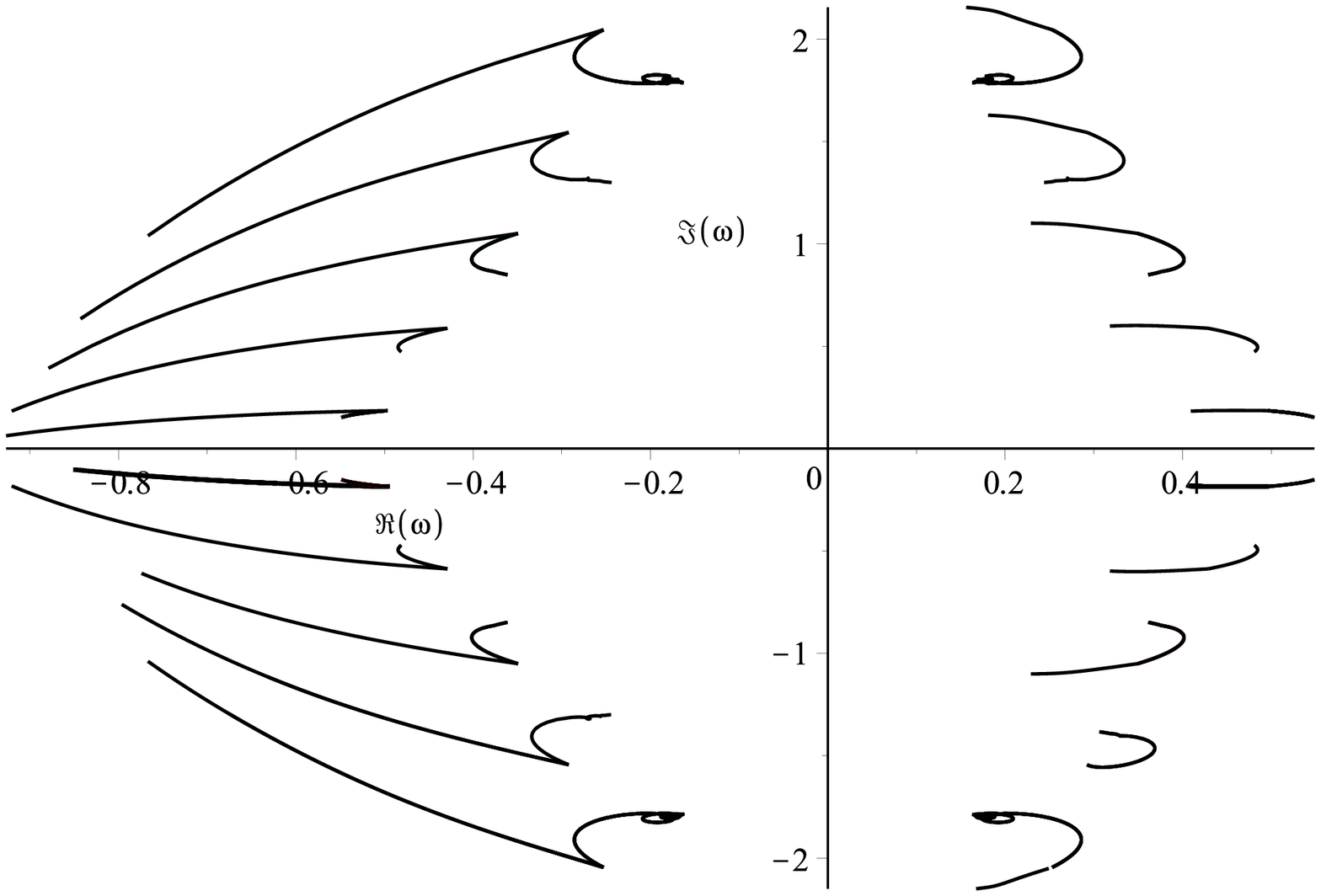}}
\subfigure{\includegraphics[width=117px,height=110px]{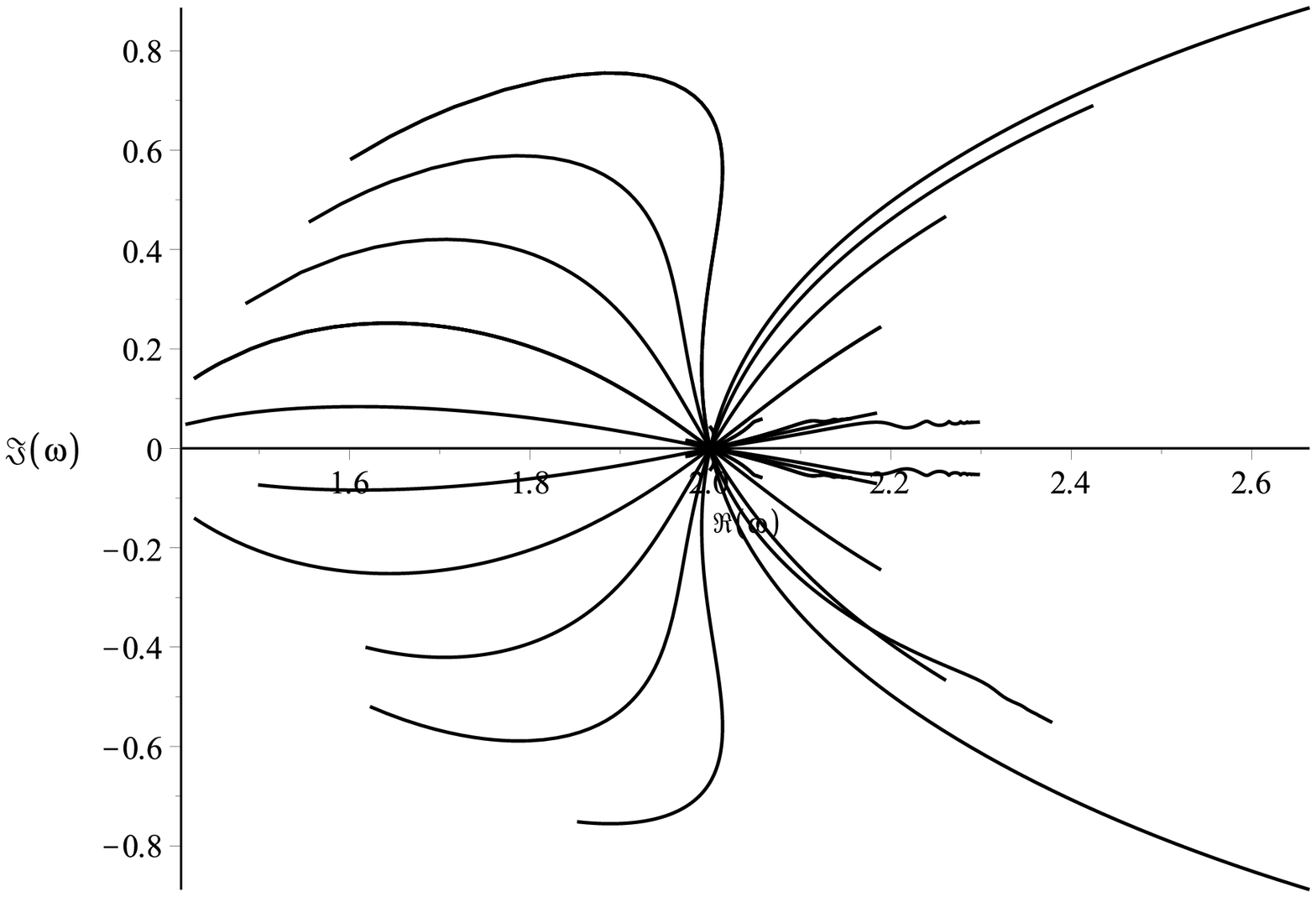}}
\caption{A complex plot of all the $\omega_{m,n}(a)$ and $E_{m,n}(a)$ obtained for $a=[0,M)$ for $m=0,1, l=1$ $n=0..4$}
\label{m01_all}
\end{figure}

On Fig. \ref{n3}, one can see an example of the loops characteristic for modes with $m=0, n\ge 3$. On Figs. \ref{m0_all_o} and \ref{m01_all}, one can see all the results plotted together.

\begin{figure}[!htb]
\centering
\subfigure{\includegraphics[width=117px,height=110px]{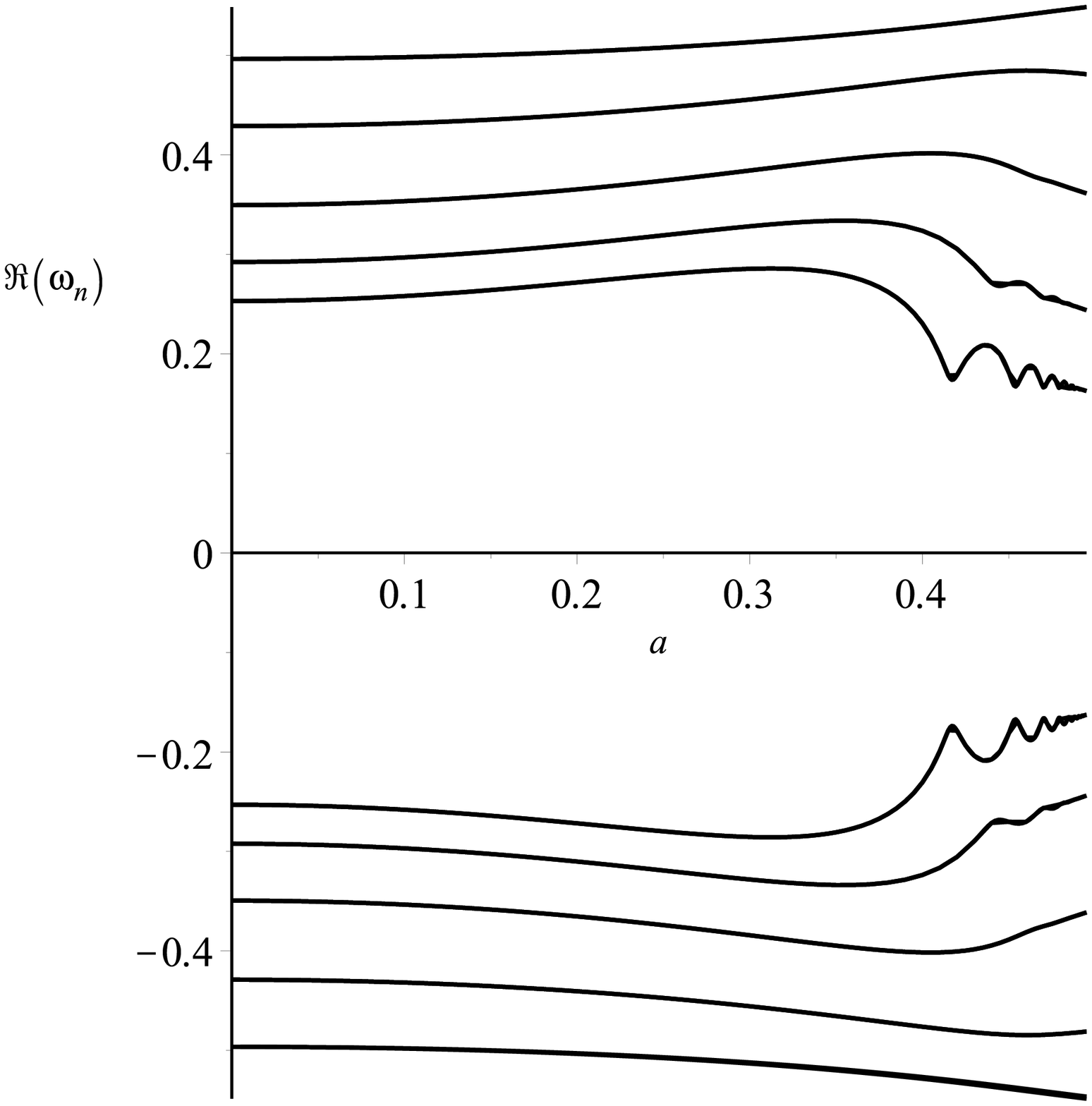}}
\subfigure{\includegraphics[width=117px,height=110px]{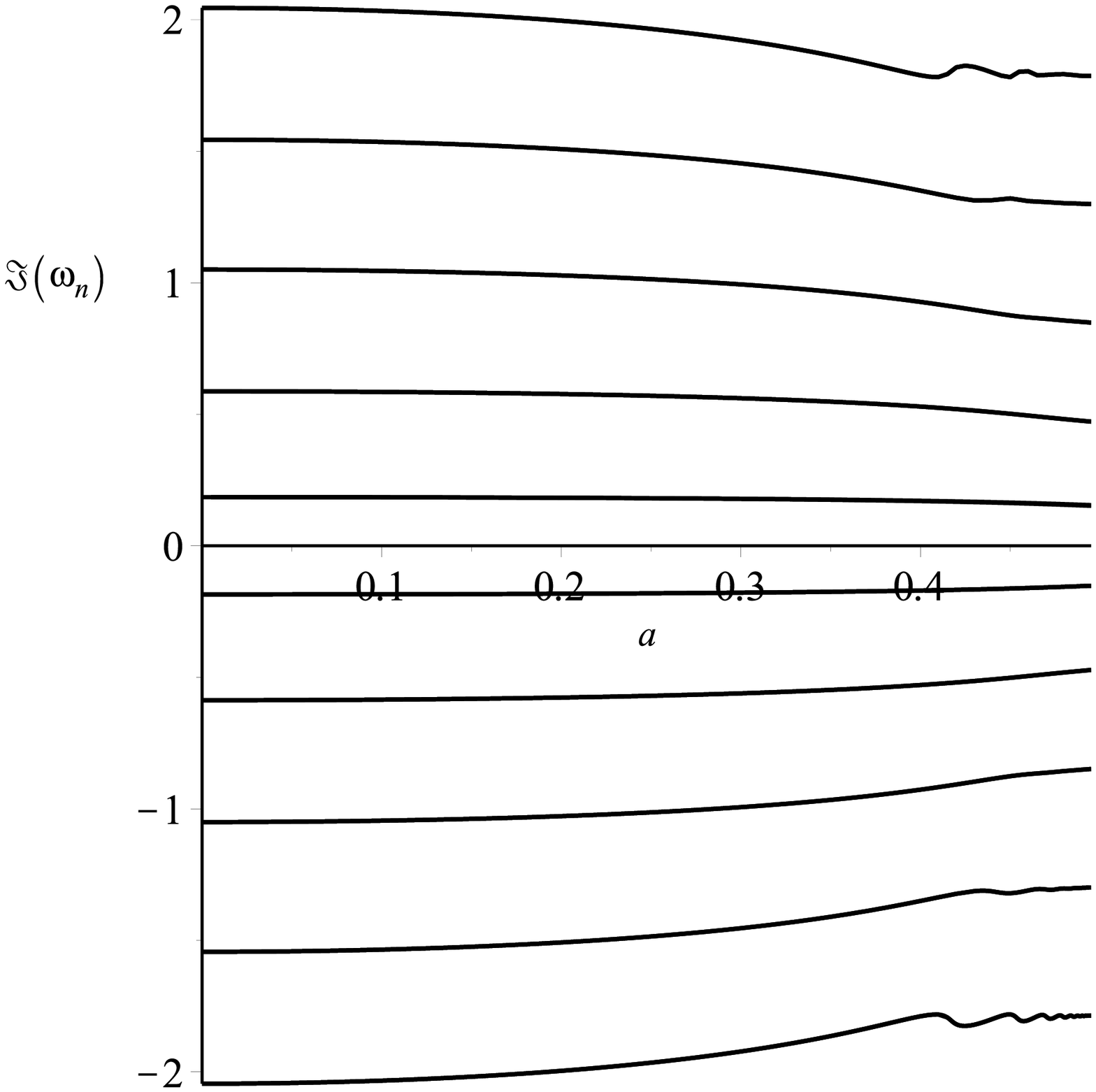}}
\subfigure{\includegraphics[width=117px,height=110px]{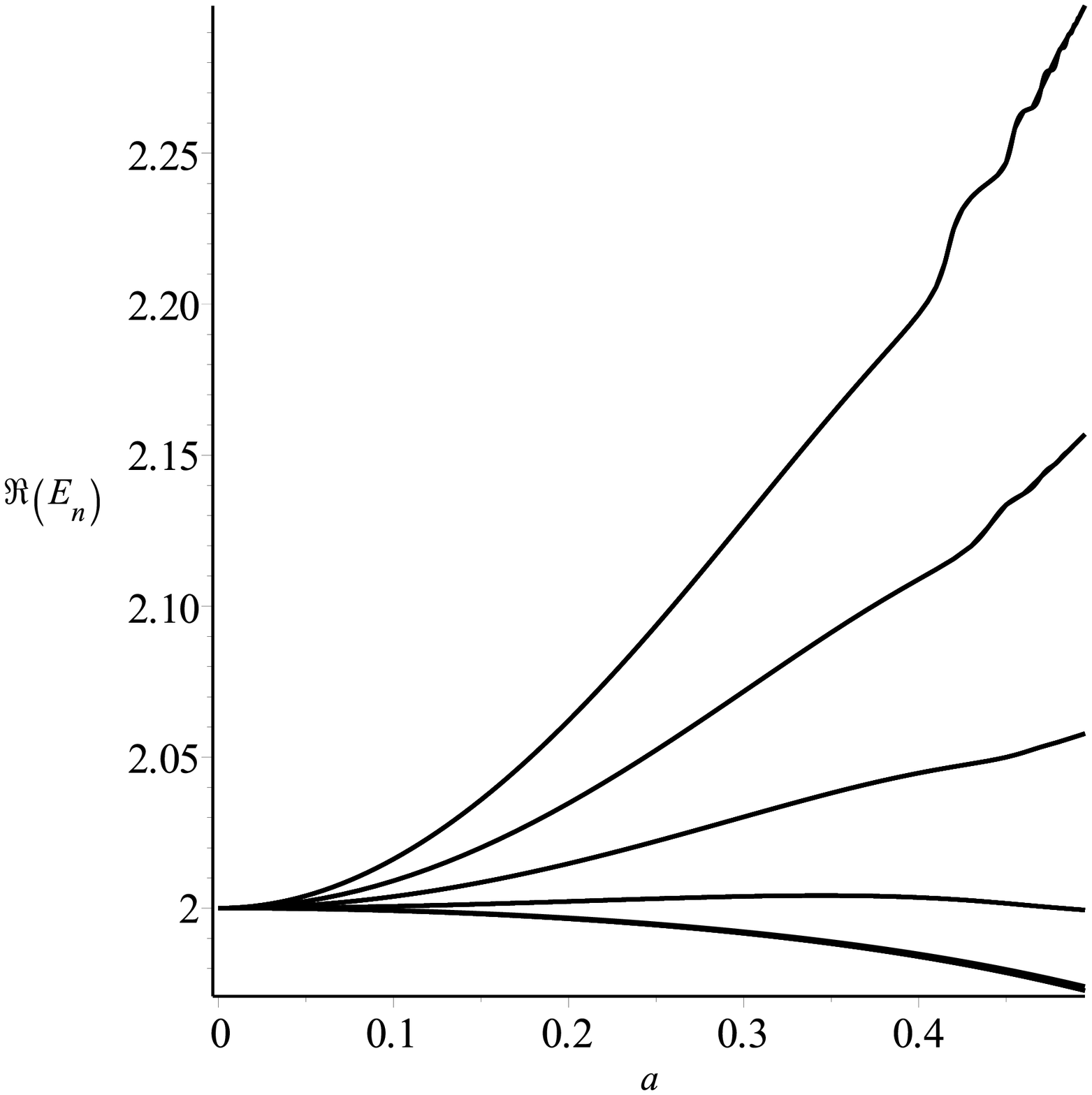}}
\subfigure{\includegraphics[width=117px,height=110px]{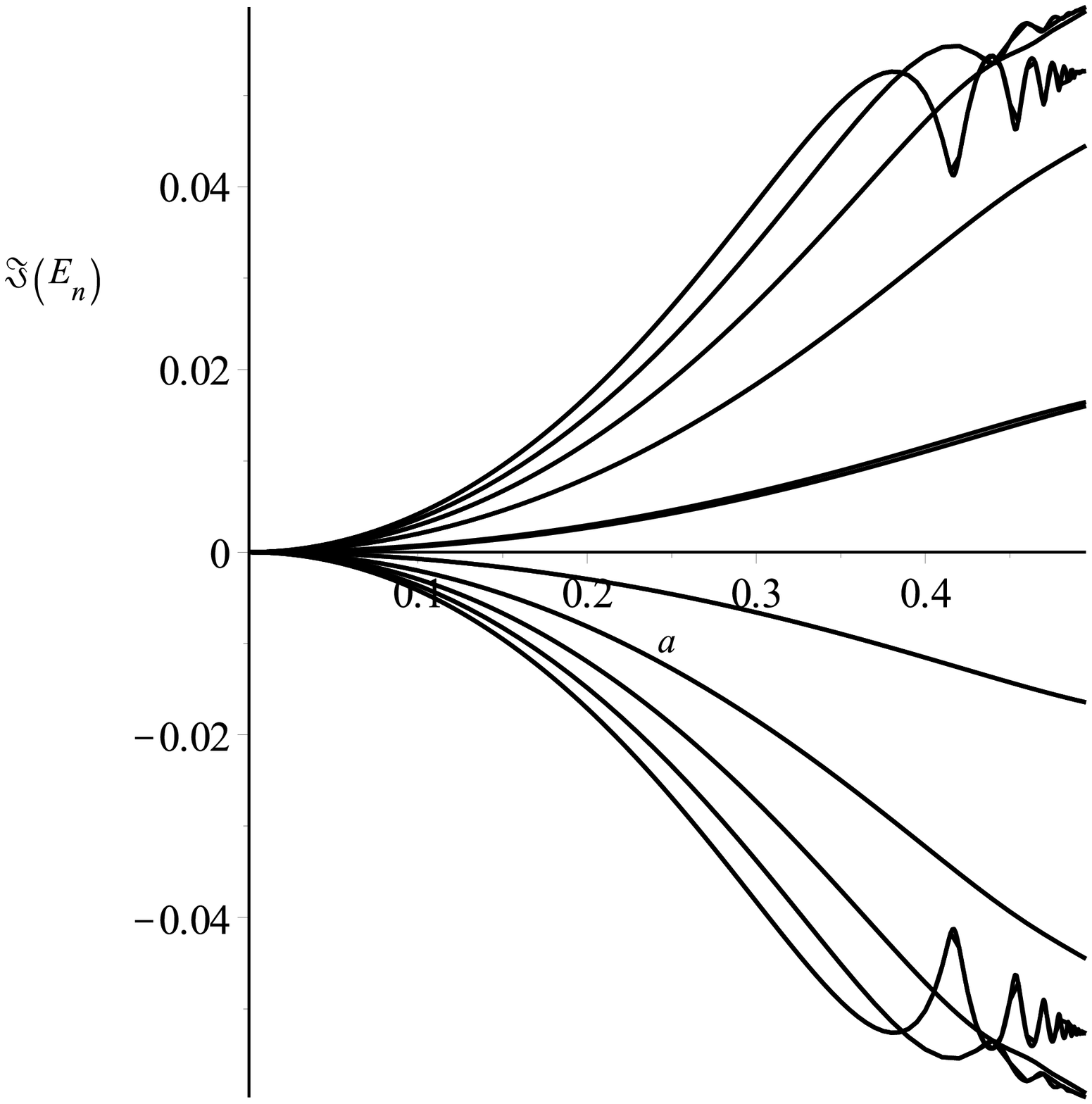}}
\caption{On the plots the real and the imaginary parts of $\omega_{0,n}(a)$ and $E_{0,n}(a)$ for $a=[0,M)$ for the modes $n=0..4$.}
\label{m1n7_21}
\end{figure}

\begin{figure}[!htbp]
\centering
\subfigure{\includegraphics[width=117px,height=110px]{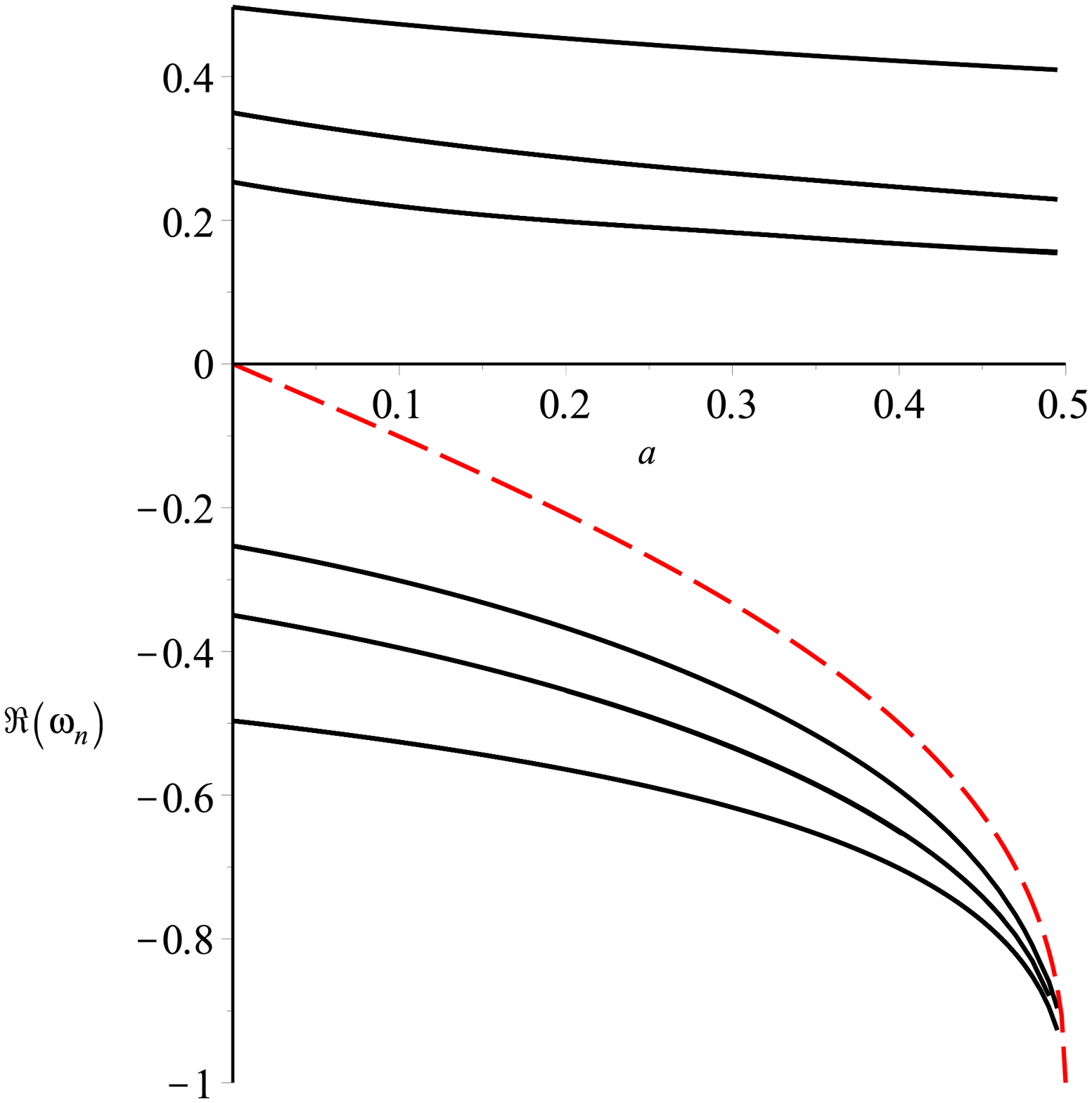}}
\subfigure{\includegraphics[width=117px,height=110px]{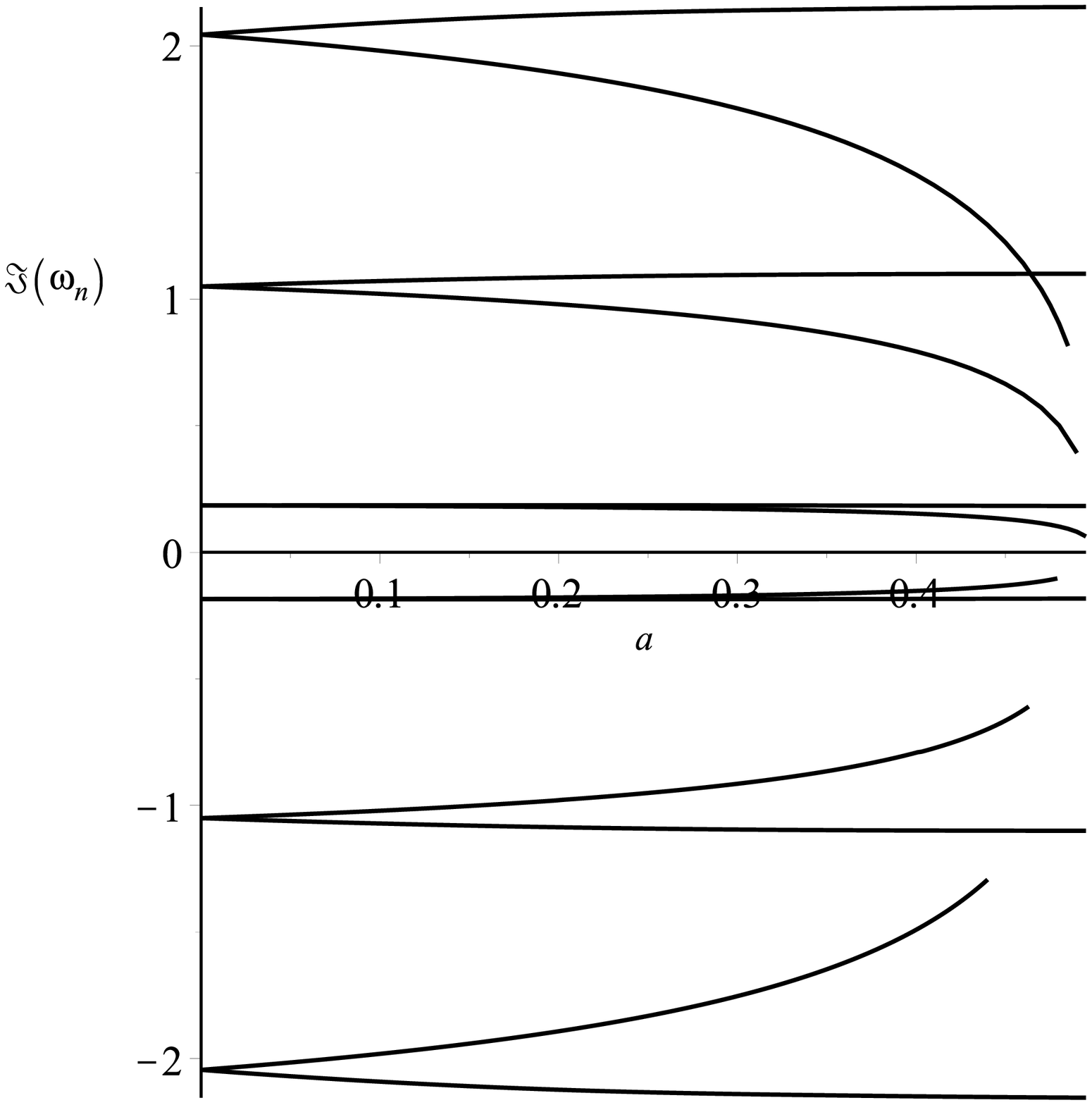}}
\vspace{-0.5cm}
\caption{On the plots: with black solid lines -- $\Re(\omega_{1,n})(a)$ and $\Im(\omega_{1,n})(a)$ for $a=[0,M)$, $n=0..2$ and with red dashed line $-m\frac{a}{2Mr_+}$ for $m=1$}
\label{m1_cr}
\end{figure}

As discussed in previous sections, only frequencies for which $\Re(\omega)\!\not\in \!(0,\!-\!m\frac{a}{2Mr_+})$ and roots of $R_2(r)$ correspond to black hole boundary conditions. From Fig. \ref{m1_cr} a), it is clear that the so obtained QNM spectrum obeys this condition. A deviation from this condition was observed in \cite{spectra}, where some of the frequencies describing primary jets crossed the line defined by $\!-m\frac{a}{2Mr_+}$, thus corresponding to a white hole solution. For the QNM spectrum, however, this is not the case and the spectrum corresponds to perturbation of a black hole. Similarly, for the QBM spectrum in quadrant III, because one is working with the roots of $R_1(r)$, for $\omega_n<-m\frac{a}{2Mr_+}$, the frequencies correspond to a white hole solution, which combined with the boundary condition at infinity, i.e. $\sin(\arg(\omega)+\arg(r))>0$ leads to QBM. Importantly, rotation doesn't change the type of the spectrum. 

{\bf Near-extremal regime:} From the same figure, one can see in the negative sector of the plot (i.e. frequencies with $\Re(\omega_n)<0$), that the real parts of the QNMs for increasing $n$ seem to tend to the line $-m\frac{a}{2Mr_+}$, confirming the observed in \cite{high} relation $\Re(\omega)=-m$ for $a\to M$. For the frequencies in the positive sector ($\Re(\omega_n)>0$), up to $a=0.4997$, the frequencies for $m=1$ do not seem to tend to a single limit. 

\begin{figure}[!htbp]
\vspace{-0cm}
\centering
\subfigure{\includegraphics[width=117px,height=110px]{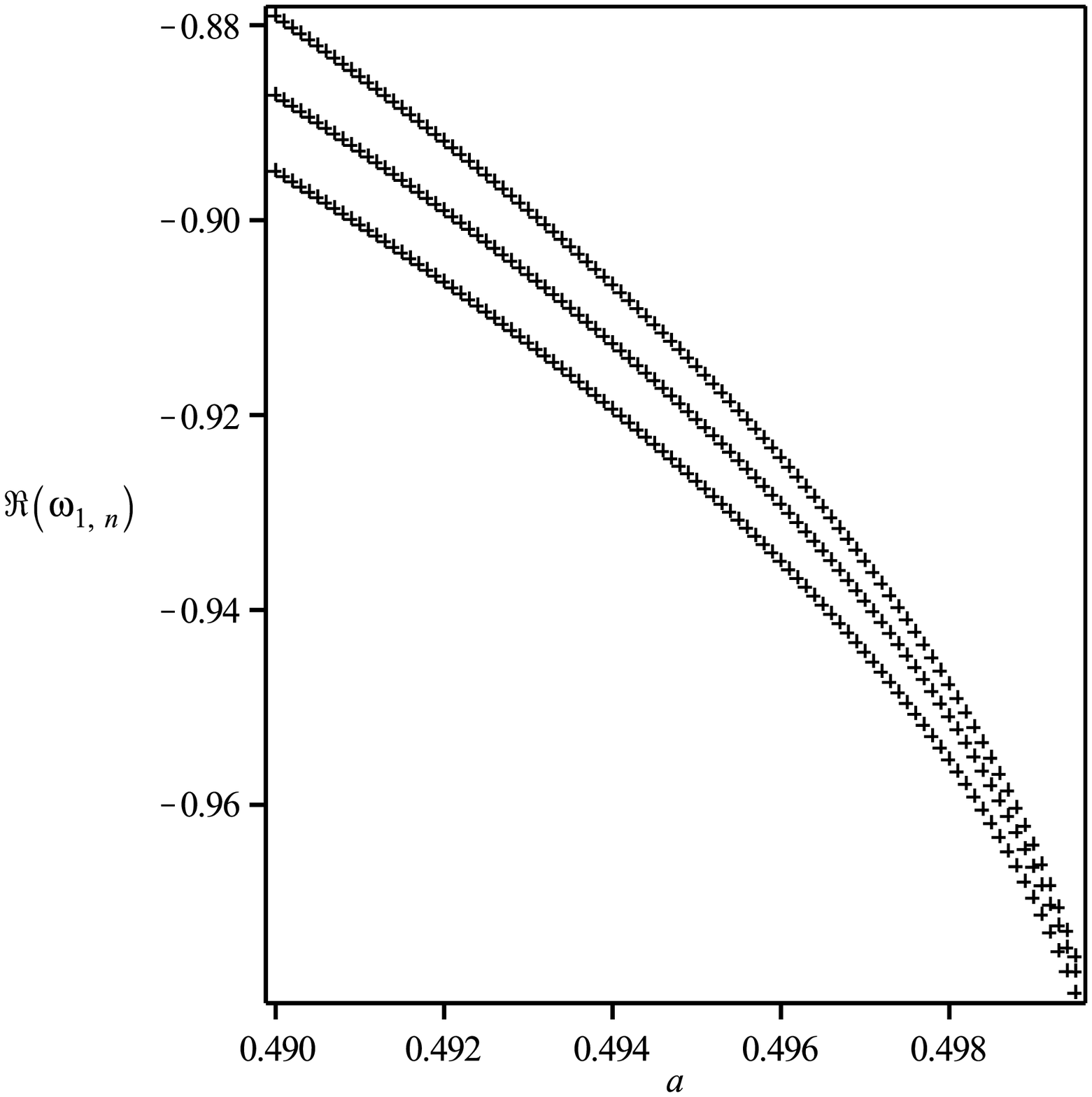}}
\subfigure{\includegraphics[width=117px,height=110px]{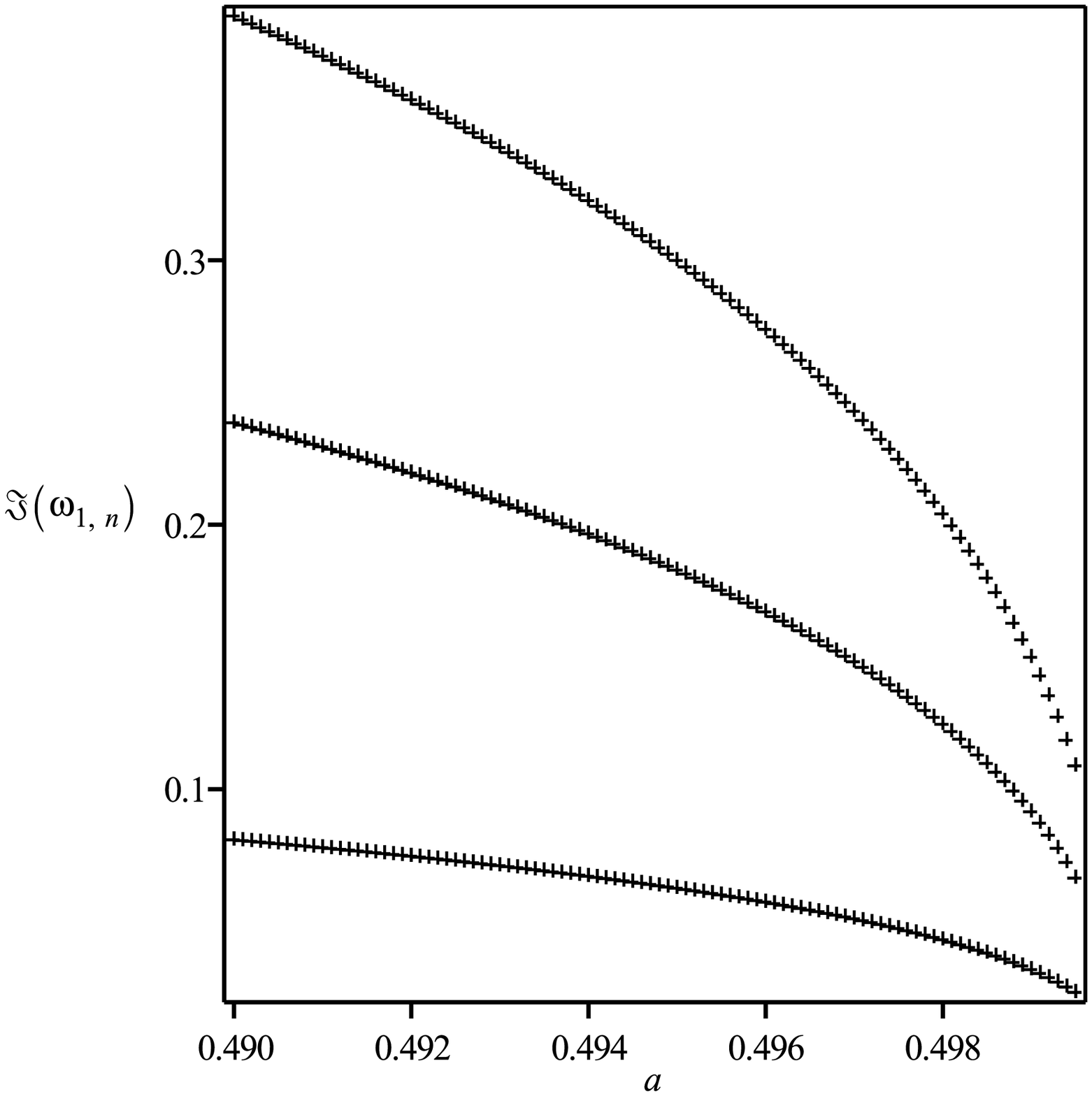}}
\vspace{-0.5cm}	
\caption{On the plot $\Re(\omega_{1,n})(a)$ and $\Im(\omega_{1,n})(a)$ for $a=[0.49,0.4995]$ for the modes $n=0,1,2$, $m=1$}
\label{aM}
\end{figure}

Obtaining the modes in the limit $a\approx M$ could be of serious interest, if one is to compare the EM QNMs with the spectra obtained from astrophysical objects but it is also technically challenging. This happens because for $a=M$ the TRE changes its type and near this limit the confluent Heun function becomes numerically unstable since the analytical solutions for $a=M$ are of the biconfluent Heun type. Due to this, the examination of the limit $a\to M$  for modes with high $n$ is very difficult with current numerical realization of that function. For the lowest modes, however, the function is stable enough in the interval $a\in[0.49,0.49999]$ and the results of the numerical experiment for $m=1$ are plotted on Fig. \ref{aM}. As expected, for $n=0$, for $a>0.91M$ the imaginary part of the frequency quickly tends to zero, thus proving that for extremal objects, the perturbations damp very slowly. The other two modes also seem to tend to zero, although more slowly than $n=0$. In physical units, the difference between the 3 modes for $a=0.4995$ is only 6Hz ($\omega_{0,1}\approx 1.582kHz$), but the damping times of the first mode is approximately 4.86 times bigger than that of the third and is $t^{damp}_{0,1} \approx 4.2ms$ for KBH with mass $M=10 M_\odot$. Note that the milliseconds time-scales correspond to the usual lifetime of central engines in numerical simulations, once again confirming the relevance of QNMs. 

As a test of the numerical stability of our method in the near-extremal case, we compare our numerical results with analytical formulas Eqs. 11 and 12 in \cite{EBH} where this regime has been theoretically studied. We considered the cases m=1, l=1, n=0..2 and m=2, l=2, n=0 in the extremal limit $a/M\in[0.995..0.99998]$. It was found that the equations describe well the numerical results, with Eq. 11 fitting better the imaginary parts ($\Im(\omega_{theor})/\Im(\omega_{num}) \sim 1\pm 0.05$), while Eq. 12 fits better the real parts ($\Re(\omega_{theor})/\Re(\omega_{num})\sim 1\pm 0.01$). The best value for the parameter ''$\delta$`` for $m=1,2$ is $\delta=\!-\!1/2$ (for $m=1$ also $\delta\!=\!-\!1/2\!+\!i/10$). 
The good agreement of Eq. 12 with the numerical data is somewhat unexpected, since in all test cases we have equatorial modes ($m=l$), where Eq. 11 should work better. While this behavior needs to be investigated further, the fact that the analytical formulas approximate the numerical results in the near-extremal regime well serves as an independent confirmation of the viability of the method.

\subsection{Algebraically special modes, branch cuts and spurious modes}

The algebraically special (AS) modes are obtained from the condition that the Teukolsky-Starobinsky constant vanishes (\cite{QNM0}) and they correspond to the so called total transmission modes (TTM) -- modes moving only in one direction: to the right or to the left. In the case of gravitational perturbations ($s=-2$) from non-rotating BH, because the $9^{th}$ QNM coincides approximately with the theoretically expected purely imaginary AS mode, there were speculations that the two modes coincide (see \cite{high} for a review, and also \cite{special2,AS}). A study of this mode in the case of gravitational perturbations of KBH showed numerical peculiarities as the ``doublet'' emerging from  the ``AS mode'' for $m>0$ (\cite{high}). For the non-rotating gravitational case, Maassen van den Brink \cite{special2,AS} found that the peculiarities of the $9^{th}$ mode are due to the branch cut in the asymptotics of the Regge-Wheeler potential, which the method of continued fractions is not adapted to handle. This result was confirmed by the use of the $\epsilon$-method in our previous work \cite{arxiv3}, where the AS character of the $9^{th}$ mode was disproved.

For electromagnetic perturbations, the algebraically special modes have not been discussed much, because in the limit $a\to 0$, the Teukolsky-Starobinsky constant do not vanish for purely imaginary modes (in fact, for $a=0$ the Teukolsky-Starobinsky constant does not depend on $\omega$ at all, see Eq. (60) \cite{QNM0} p.392) and there appear to be no correlation between TTM and QNM modes \cite{AS1}.

Because of the use of the analytical solutions, instead of approximate method such as the continued fraction, we are able to study the behavior of the modes with respect to proximity of a branch cut. As discussed in Section 4, we can use the epsilon-method to rotate the branch cut in the complex $\omega$-plane. This is because thanks to the relation coming from the QNM infinity boundary condition  $\arg(r)\!+\!\arg(\omega)\!=\!\frac{3+\epsilon}{2}\!\pi$ (similarly for QBM) we can switch between $\omega$ and $r$. Even more importantly, because of the use of analytical solutions, we have a clear understanding how to differentiate between physical modes such as QNM and QBM and the so-called spurious modes.  

\begin{figure}[!htbp]
\vspace{-0cm}
\centering
\subfigure{\includegraphics[width=117px,height=110px]{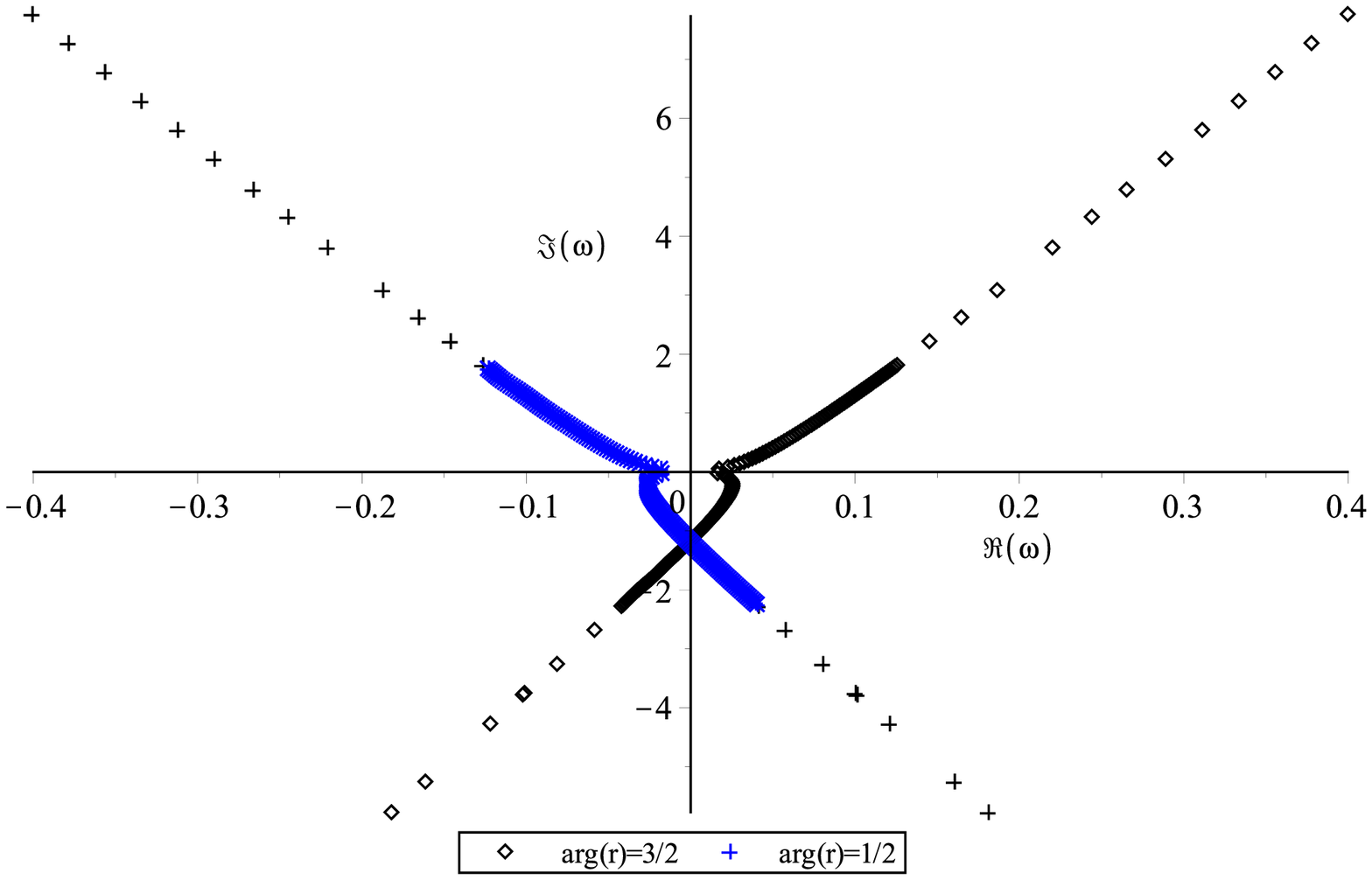}}
\subfigure{\includegraphics[width=117px,height=110px]{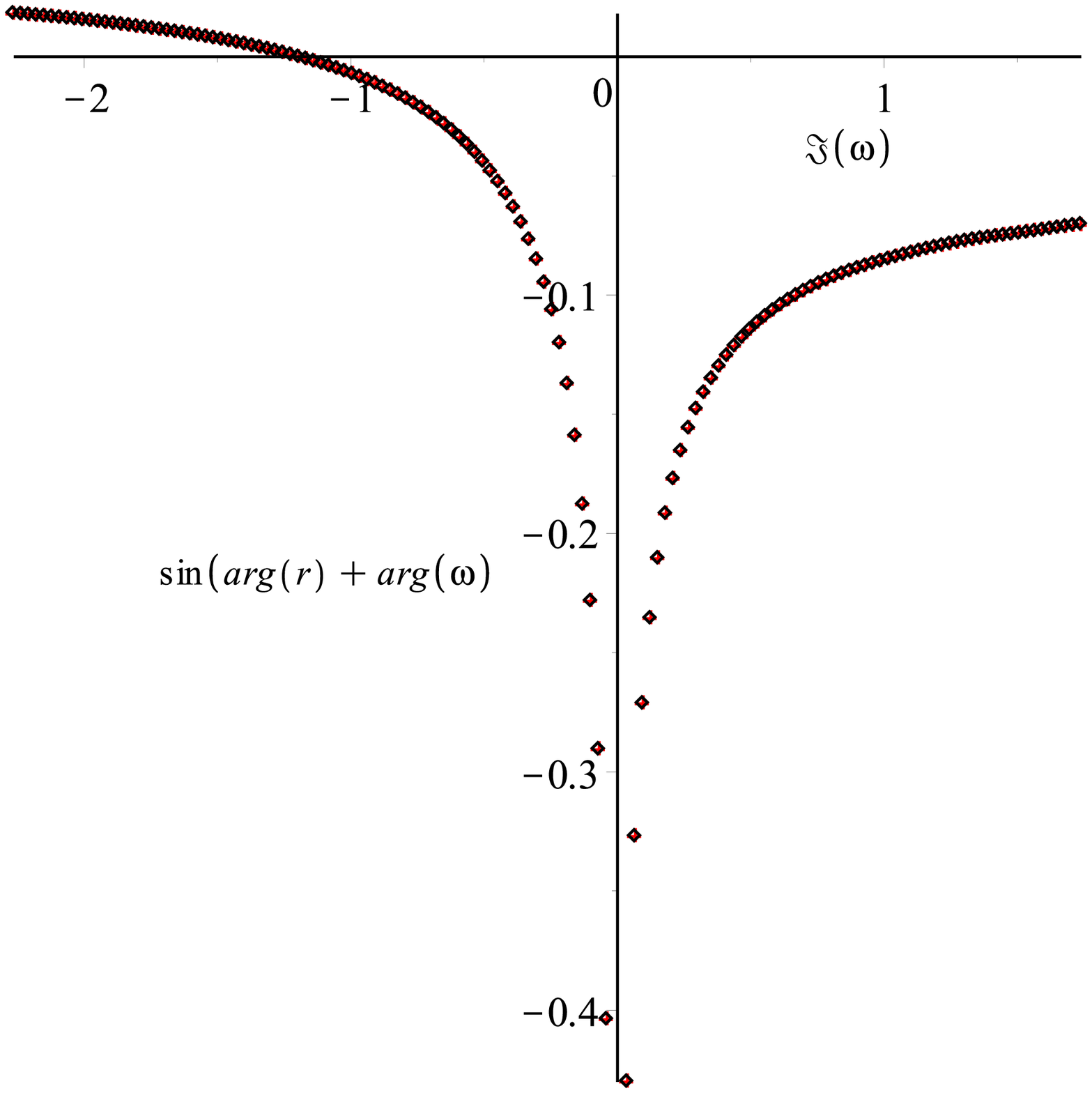}}
\vspace{-0.5cm}	
\caption{a) The spurious modes found as roots of the system. b) the boundary condition for them. Clearly, those modes fulfill both the QNM and the QBM condition for certain $n$. }
\label{spurious}
\end{figure}

An example of such a spurious additional spectrum can be seen on Fig. \ref{spurious}. In this case, the modes seem to fulfill the QNM boundary condition for certain $n$ and the QBM boundary condition for other $n$. This makes understanding the nature of those modes difficult. An additional test of their numerical stability with respect to changes in $|r|$ shows that those modes are unstable and they basically decrease with the increase of the actual infinity. Because  $\omega \nsim r$ is underlying assumption of the problem, we discard those modes as unphysical. Such spurious modes have been reported in \cite{high} without further explanation of their nature. In our case, those are modes which are solutions of the spectral system, but which contradict the derivation of the TRE and TAE. 

Additional spurious modes can be observed outside of the stability intervals on $\epsilon$ for the QNMs/QBMs. For example, for $n=0, m=0, a=0$, one has a stable mode with precision of 20 digits in the interval $\epsilon=-0.5..0.35$. For $\epsilon>0.35$ the mode starts increasing its real part until it reaches a value for which the boundary condition at infinity is no longer satisfied. We consider those modes to be a product of numerical instability due to inadequate choice of the direction of the steepest descent. Another possible reason for such dramatic changes in the numerical stability of the mode with respect to $\epsilon$ are that the branch cut in the radial function is moved by the parameter $\epsilon$, i.e. they are due to the complex character of the used analytical functions (the confluent Heun functions) in the vicinity of the irregular singular point $r=\infty$ in the complex $r$-plane. Understanding better the theory of the confluent Heun function is critical for the complete understanding of the numerical results. 

\section{Discussions and conclusion}
From the recent developments in the field of gravitational waves detection, it is clear that finding the EM counterpart to those events can prove to be very useful. In this case, it is needed a better understanding of the fundamental physics of quasinormal ringing. In this paper, we offered a new approach to finding the QNMs for the KBH, based on directly solving the system obtained by the analytical solutions of the TRE and TAE in terms of the confluent Heun function. This approach has the advantage of being more traditional  (i.e. imposing directly the corresponding boundary conditions on the exact analytical solutions of the problem) and hence it should allow better understanding of the peculiar properties of the EM QNMs and the physics they imply.

It was shown that using this approach one can reproduce the frequencies already obtained by other authors, but without relying on approximate methods. Particularly important is the ability to impose the boundary condition {\em directly} on the solutions of the differential equations. We require the standard regularity condition on the TAE and explore in detail the radial boundary condition (the BHBC).
We then can solve the system for each of the radial equations, solution of the Teukolsky Master Equation and find both quasinormal and quasibound modes, and additional spurious spectra. By tracking the boundary condition at infinity we are able to work with all those modes at the same time and to obtain their spectra with and without rotation. Such a result shows the advantage of our novel method over more traditional methods. It also raises the question of the theory of the Heun functions as a critical part of understanding the numerical results.

The key results of the article are as follow:

1. High-precision reproduction of known QNM results and study of the numerical stability of the modes in the radial complex plane.

2. New QBM spectrum, obtained for the inverse boundary conditions, its numerical stability also studied.

3. An additional spurious spectrum found to contradict assumptions of the problem and thus classified as numerical artifact. 

From our result, the numerical stability of the modes in the r-plane turned out to be a powerful tool for understanding the results. Although obvious in our formulation of the problem, such a study is impossible to perform with the method of continued fraction (at least not directly) because of the missing r-variable. Its usefulness, however, is clear, when one considers the fact that the spurious modes evolve with rotation and can be seen even in the naked singularity regime, where the QNMs and QBMs disappear. 

With respect to the QBM spectrum, there is a need of clarification. The Boyer-Lindquist coordinates for the Kerr metric have the following ranges $0\le r \le \infty, 0\le \theta \le \pi, 0 \le \phi \le 2\pi, -\infty <t<\infty$. The stability of the object with respect to perturbations depend on the standard substitution $\Psi=e^{i(\omega t+m\phi)}S(\theta)R(r)$. Clearly, the stability condition $\Im(\omega)>0$ will mean the perturbation will damp with time only as long as $t>0$. Since the range of the t-variable includes also negative values of the time, in those cases, we will have exponentially growing perturbations for the same condition. For the QBM spectra, however, since $\Im(\omega)<0$ the perturbations will damp with time for $t<0$. Also, because of the symmetries of the Kerr metric, it is known that the metric is invariant under the changes $t\to-t, \phi \to -\phi$, meaning that the inversion of time can be considered as inversion of the direction of rotation. Therefore, one can say that the new modes are not unphysical if considered in the correct range for the time-variable. A question of interpretation is whether we will consider the boundary conditions as unstable quasibound (i.e. inverse to QNM) modes for $t>0$ or as stable QNMs for $t<0$. 

As a conclusion, the new approach proved that using the analytical solutions of the TRE and TAE, along with the direct imposing of boundary conditions, allowed for better understanding of the problem and for obtaining with high precision both the well-known spectrum and its symmetric but essentially new spectrum. 

\section{Acknowledgments}
The authors would like to thank Prof. E. Berti for discussion of the numerical values of the EM QNM frequencies  obtained within the Leaver method, which was important for the comparison of our method with this already well-established one. 

The authors would like to thank Dr. Edgardo Cheb-Terrab for useful discussions of the algorithms evaluating the Heun functions in \textsc{maple} and for continuing the
improvement of those algorithms.

The authors would like to thank the anonymous reviewer who drew our attention also to the work of Shahar Hod, based on analytical calculations. The comparison of our numerical results with that work increased our confidence in the validity of the results. 

This article was supported by the Foundation "Theoretical and
Computational Physics and Astrophysics", by the Bulgarian National Scientific Fund
under contracts DO-1-872, DO-1-895, DO-02-136, and Sofia University Scientific Fund, contract 185/26.04.2010, Grants of the Bulgarian Nuclear
Regulatory Agency for 2013, 2014 and 2015.

\section{Author Contributions}
P.F. posed the problem of evaluation of the EM QNMs of rotating BHs as a continuation of previous studies of the applications of the confluent Heun functions in astrophysics. He proposed the epsilon method and supervised the project.

D.S. is responsible for the numerical results, their analysis and the plots presented here.

Both authors discussed the results at all stages. The manuscript was prepared by D.S. and edited by P.F..

\makeatletter
\let\clear@thebibliography@page=\relax
\makeatother

\end{document}